\documentclass[superscriptaddress,twocolumn]{revtex4-1}

\usepackage[a4paper, total={7in, 10in}]{geometry}
\usepackage{amsmath,amssymb,graphicx,color,amsfonts,hyperref}

\renewcommand{\eqref}[1]{{Eq.~(\ref{#1})}}

\begin{document}
\title{Dissipative light  bullets in a doped and weakly nonlocal optical fiber}
\author{Ghislaine Flore Kabadiang Ngon}
	\email {ngonghislaine@yahoo.fr}       
\affiliation{Laboratory of Mechanics, Department of Physics, Faculty of Science, University of Yaound\'{e} I,
P.O. Box 812, Yaound\'{e}, Cameroon}
\author{Conrad Bertrand Tabi}
	\email {tabic@biust.ac.bw}
	\affiliation{Department of Physics and Astronomy, Botswana International University of Science and Technology, Private Mail Bag 16 Palapye, Botswana }

\author{Timol\'{e}on  Cr\'{e}pin Kofan\'e}
\email {tckofane@gmail.com}
	%\homepage       {http://www.biust.ac.bw}
	\affiliation{Department of Physics and Astronomy, Botswana International University of Science and Technology, Private Mail Bag 16 Palapye, Botswana }
	\affiliation{Laboratory of Mechanics, Department of Physics, Faculty of Science, University of Yaound\'{e} I,
P.O. Box 812, Yaound\'{e}, Cameroon}

\date{\today}
	
\begin{abstract}
The letter introduces an extended (3+1)-dimensional [(3+1)D]  nonlocal cubic complex Ginzburg-Landau equation describing the dynamics of dissipative light bullets in optical fiber amplifiers under the interplay between dopants and a spatially nonlocal nonlinear response. The model equation includes the effects of fiber dispersion, linear gain, nonlinear loss, fiber nonlinearity, atomic detuning, linear and nonlinear diffractive transverse effects, and nonlocal nonlinear response. A system of coupled ordinary differential equations for the amplitude, temporal, and spatial pulse widths and position of the pulse maximum, unequal wavefront curvatures, chirp parameters, and phase shift is derived using the variational technique. A stability criterion is established, where a domain of dissipative parameters for stable steady-state solutions is found. Direct integration of the proposed nonlocal evolution equation is performed, which allows us to investigate the evolution of the Gaussian beam along a doped nonlocal optical fiber, showing stable self-organized dissipative spatiotemporal light bullets.  
\end{abstract}
\maketitle
\section{\label{sec:1}Introduction}
Optical solitons have promising potential to become principal information carriers in telecommunication due to their capability to propagate long-distance signals without attenuation and change their shapes. One of the major goals in the field of soliton physics is the production of light fields that are localized in all three dimensions of space and time, which we will refer to as 3D spatiotemporal solitons or light bullets. This results from the simultaneous balance of diffraction and GVD by the transverse self-focusing and nonlinear phase modulation in the longitudinal direction, respectively~\cite{MalomedJOB2005}.

Considerable attention is being paid to theoretically and experimentally analyzing spatial optical solitons' dynamics in material with local nonlinear and nonlocal responses~\cite{KivsharBook2003,SuterPRA1993,RotschildPRL2005}. Local nonlinearity of optical media is usually approximated by a local function of the light intensity assuming the refractive index change at a given spatial location depends solely on the light intensity at the same location, while nonlocality means that the change of the refractive index in a particular point is determined by the light intensity not only in the same point but also in its vicinity. Recently, it has been revealed that nonlocality can provide new effects such as strong modification of modulational instability~\cite{KrolikowskiJOB2004,TagwoOptik2017}, suppression of beam collapse~\cite{O.BangPRE2002}, dramatic change of the soliton interaction~\cite{PecciantiOpLet2002,NikolovOpLet2004}, formation of multi-soliton bound states~\cite{KartashovOpLet2005},  stabilization of spatially localized vortex solitons~\cite{DesyatnikovOpLet2005}, symmetry breaking azimuthal instability~\cite{YakimenkoPRE2005,BriedisOpEx2005} as well as stabilization of different nonlinear structures such as ringlike clusters of many solitons~\cite{DesyatnikovPRL2002}  and modulated localized vortex beams or azimuthons~\cite{DesyatnikovPRL2005}. 
		
In long-distance soliton propagation, the energy of the soliton decreases because of fiber losses. This would produce soliton broadening because a reduced peak power weakens the SPM effect necessary to counteract the effect of GVD. Therefore, soliton must be amplified periodically using either the lumped or the distributed amplification scheme to overcome the effect of fiber losses.  It was demonstrated that by doping the optical fiber with rare earth ions, such as Neodymium, Erbium, Praseodymium, and Ytterbium ions, just to name a few, energy gains were made in addition to the optical fiber. Thus, the light signal could be amplified each time it weakened, and the loss of information could be avoided~\cite{DesurvireOpLet1987}. On the other hand, there are interesting studies on the pulse propagation problem in doped fiber amplifiers within the rate equation approximation, the governing equation being the cubic complex Ginzburg-Landau (CGL) equation and its variants~\cite{AgrawalPRA1991,ManousakisOpCom2002,FewoOpCom2008,FewoJPSJ2008,TiofackPRE2009,DingJOSAB2009}. Dissipative solitons can be propagated in such active fiber over long distances. In some previous studies~\cite{SoljacicPRL2001,TowersJOSAB2002,MatuszewskiPRE2004,HePRE2006,KamagatePRE2009,DjokoOpCom2018,MalomedJOB2016}, multidimensional spatiotemporal optical solitons, i.e., both (2+1)-dimensional [(2+1)D] and (3+1)-dimensional [(3+1)D] dissipative optical bullets were considered taking the (2+1)D and (3+1)D family of CGL equations. Stable spatiotemporal dissipative solitons have been reported. Among them are stationary stable and pulsating solutions, double, quadruple, sixfold, eightfold, and tenfold bullet complexes, self-trapped necklace-ring, ring-vortex solitons, uniform ring beams, spherical and rhombic distributions of light bullets, fundamental and cluster solitons, respectively. Recently, the stability diagram obtained from the Lenz transformation and the linear stability analysis has revealed that higher values of the quintic nonlocality contribute to reducing the modulational instability (MI) in weakly cubic-quintic nonlocal nonlinear media~\cite{ZangaCNSNS2020}.  Very recently, it has been shown that instability regions from the pure quartic MI gain in weakly nonlocal birefringent fibers are more expanded due to nonlocality, which was confirmed via direct numerical simulations showing the emergence of Akhmediev breathers~\cite{TabiOpLet2022,TiofackJOpt2023}. For a nonlinear saturable media with competing nonlocal nonlinearity, it was reported that the quenching effect of the nonlocal nonlinearity on the MI is corrected, especially when the saturable index and the nonlocality range are well-balanced~\cite{TabiPRE2022}.
		
The main purpose of the present work is to investigate (3+1)D dissipative light bullets in fiber amplifiers under the interplay between dopant and a spatially nonlocal nonlinear response which, to the best of our knowledge, has not yet been proposed in the literature. The dopant is modelled as a two-level system whose dynamic response is governed by the population and dipole relaxation times.  For incident optical of width such that, in cases of weak nonlocality and in the paraxial wave approximation, the (3+1)D cubic complex Ginzburg-Landau equation is derived, which includes the effects of fiber dispersion, linear gain, nonlinear loss, fiber nonlinearity, atomic detuning, linear and nonlinear diffractive transverse effects, and nonlocal nonlinear response. We mainly focus on the localized Gaussian solution in the form of three-dimensional traveling waves. Our main purpose is to assess the role played by a weak spatial nonlocality term in the shape formation of a dissipative light bullet. A system of eight coupled first-order differential equations of the solution's parameters of interest is derived on the basis of variational equations resulting from the Euler-Lagrange equations.
		
The remaining part of this paper is organized as follows. In Sec.~\ref{sec2}, we derive the (3+1)D nonlocal cubic complex Ginzburg-Landau (CGL) equation governing the dynamics of the dissipative light bullets in a doped and weakly nonlocal nonlinear medium. In Sec.~\ref{sec3}, the dynamic characteristics of the dissipative light bullets, such as the amplitude, temporal and spatial pulse widths, the position of the pulse maximum, unequal wavefront curvatures, chirp parameters, and the phase shift in specially designed media, are studied using the variational technique. In Sec.~\ref{sec4},  a stability criterion for steady-state solutions of the (3+1)D nonlocal cubic CGL equation is established, fixing a domain of dissipative light bullets parameters. In Sec.~\ref{sec5}, the direct integration of the  (3+1)D nonlocal cubic CGL equation with the Runge-Kutta and the split-step Fourier methods have been carried out, showing stable self-organized dissipative spatiotemporal light bullets. Some concluding remarks are given in Sec.~\ref{sec6}

\section{Derivation of the nonlinear evolution equation for electromagnetic pulse propagation in doped and weakly nonlocal nonlinear media }\label{sec2}
		
The light pulse propagation problem in doped optical fiber can be solved by defining a complex dielectric constant as follows~\cite{AgrawalPRA1991}:
\begin{equation}\label{eq1}
\epsilon(\omega)= n_{f}^{2}+2in_{f}\frac{\alpha_f}{k_{0}}+\chi_{a}(\omega),
\end{equation}
where $\omega$ is the optical frequency, with $\alpha_f$ being the fiber loss defined by:
\begin{equation}\label{eq2}
\alpha_f=\frac{10}{L}\log\frac{P_{out}}{P_{in}} {\rm (dB/km)},
\end{equation}
where $P_{in}$ is the input light pulse power, $ P_{out}$ is the output light pulse power and $\textit{L}$, the length of the optical fiber. $k_{0}$ is the wavenumber given by
$k_{0}={\omega_{0}}/{c}$, with $c$ being the light speed, and $\omega_{0}$, the carrier frequency.  $\chi_{a}(\omega)$ is the atomic susceptibility governing the response of the dopant in the optical fiber, which is determined by~\cite{AgrawalPRA1991}
\begin{equation}\label{eq3}
\chi_{a}(\omega)=\frac{g_{p}}{k_{0}} \frac{(\omega-\omega_{a})T_{2}-i}{1+(\omega-\omega_{a})^{2} T_{2}^{2}},
\end{equation}
with the peak gain $g_{ p} = \sigma(N_{ 2 }- N_{ 1 })$. The parameter $\sigma $ is a transition cross-section, while   $N_{ 1}$ and $N_{ 2}$ are the atomic densities for the two-level system's lower and upper energy levels. $\omega_{a}$  is the atomic resonance frequency, and $T_{ 2}$ is the relaxation time. Moreover, one can expand the function $\chi_{ a} (\omega)$ in Taylor's series up to the second order in the vicinity of the carrier frequency $\omega_{ 0}$ and find~\cite{AgrawalPRA1991}
\begin{multline}\label{eq4}
\chi_{a}(\omega)=\frac{g_{p}}{k_{0}}\left[ \frac{\delta-i}{1+\delta^{2}}+\frac{1-\delta^{2}+2i\delta}{(1+\delta^{2})^{2}}(\omega-\omega_{0})T_{2}\right.\\ \left.  
+\frac{\delta(\delta^{2}-3)+i(1-3\delta^{2})}{(1+\delta^{2})^{3}}(\omega-\omega_{0})^{2}T_{2}^{2}\right] ,
\end{multline}
where $\delta$ = $(\omega_{ 0} - \omega_{ a} )T _{2}$  is the detuning parameter. The term $n_{ f}$,  in Eq.(\ref{eq1}), is the refractive index of the fiber, including linear, nonlinear, doping, and spatial-nonlocality phenomena. The given expression of $n_{ f}$  is~\cite{BezuhanovPRA2008}
\begin{equation}\label{eq5}
n_{f}=n_{0}(\omega)+n_{2}\frac{\int R(x-x',y-y')I(x',y')dx'dy'}{\int R(x-x',y-y')dx'dy'}.
\end{equation}
Here, $n_{0}(\omega)$ is the linear refractive index, $n_{2}$ represents the nonlinear change in the refractive index, $R(x-x',y-y')$ is the nonlocal response function which determines the spatial extent of nonlocality. $I(x',y')$ is the nonlocal intensity which acts not only on the local points $\emph{(x,y)}$,  but also in the neighboring points and can be evaluated as~\cite{BezuhanovPRA2008}
\begin{equation}\label{eq6}
\begin{split}
I(x',y')&=I(x,y)+\frac{1}{2}\frac{\partial^{2}}{\partial x^2}I(x,y)(x-x')^2\\
&+\frac{1}{2}\frac{\partial^{2}}{\partial y^2}I(x,y)(y-y')^2\\
&+\frac{\partial^{2}}{\partial x\partial y}I(x,y)(x-x')(y-y'),
\end{split}
\end{equation}
where $I(x,y)$ is  the external action of the intensity on the points $\emph{(x,y)}$. Using this assumption, Eq. (\ref{eq5}) can be rewritten as
\begin{equation}\label{eq7}
\begin{split}
n_{f}&=n_{0}(\omega)+n_{2}|\vec{E}|^{2}+\frac{n_{2}\gamma_{xx}}{2}\frac{\partial^{2}}{\partial x^2}(|\vec{E}|^{2})\\
&+\frac{n_{2}\gamma_{yy}}{2}\frac{\partial^{2}}{\partial y^2}(|\vec{E}|^{2})
+\frac{n_{2}\gamma_{xy}}{2}\frac{\partial^{2}}{\partial x \partial y}(|\vec{E}|^{2}).
\end{split}
\end{equation}
In Eq. (\ref{eq7}) , we have considered $I(x,y)=|\vec{E}|^{2}$, where $\vec{E}$ is the electric-field vector.  The coefficients $\gamma_{xx}$, $\gamma_{yy}$ and  $\gamma_{xy}$ represent the measures of weak nonlocality degree in the transverse coordinates $\emph{x}$ and $\emph{y}$, respectively. These  nonlocality degree coefficients are determined by the  relations
\begin{equation}\label{eq8}
\begin{split}
\gamma_{\xi\xi}&=\frac{\int^\infty_{-\infty}R(\xi-\xi',\eta-\eta')(\xi-\xi')^{2}d\xi'd\eta'}{\int^\infty_{-\infty}R(\xi',\eta')d\xi'd\eta'},\\
\gamma_{\eta\eta}&=\frac{\int^\infty_{-\infty}R(\eta-\eta',\xi-\xi')(\eta-\eta')^{2}d\eta'd\xi'}{\int^\infty_{-\infty}R(\eta',\xi')d\eta'd\xi'},
\end{split}
\end{equation}
and
\begin{equation}\label{eq9}
\gamma_{\xi\eta}=\frac{\int^\infty_{-\infty}R(\xi-\xi',\eta-\eta')(\xi-\xi')(\eta-\eta')d\xi'd\eta'}{\int^\infty_{-\infty}R(\xi',\eta')d\xi'd\eta'},
\end{equation}
for $(\xi,\eta)= (\emph{x},\emph{y})$.  The term $\gamma_{\xi\eta}$ can be neglected in a suitably chosen and rotated coordinate system. Here, we consider the case of the so-called Gaussian nonlocal response functions~\cite{SkupinPRE2006}
\begin{equation}\label{eq10}
R(r-r')=\frac{1}{\pi \sigma^2}e^{-\frac{(r-r')}{\sigma^2}},
\end{equation}
with a characteristic width $\sigma$ defining the degree of nonlocality. Indeed, it has been shown that, for a Gaussian response, $\gamma_{\xi\xi}$=$w^{2}_{_{R\xi}}/2$, with $w_{_{R}}=0.1 \sigma/\sqrt{2}$. We recall that Eq. (\ref{eq6}) is justified~\cite{BezuhanovPRA2008}  when the nonlocality becomes weak, that is, $\gamma_{\xi\xi}\ll 1$. Another necessary parameter to consider, as much as the dielectric constant, in the propagation of light in a doped optical fiber, is the propagation constant around the frequency $\omega$ given as follows~\cite{AgrawalPRA1991} 
\begin{equation}\label{eq11}
\beta(\omega)= \frac{\omega}{c}\sqrt{	\epsilon(\omega)}.
\end{equation}
From Eqs. (\ref{eq1})-(\ref{eq7}),  we can rewrite the propagation constant expressed in Eq. (\ref{eq11}). Given that the term $n_{0}(\omega)\frac{\omega}{c}$ corresponds to the propagation constant of the undoped fiber denoted $	\beta_{f}(\omega)$, we then obtain the following equation		
\begin{equation}\label{eq12}
\begin{split}
\beta(\omega)&=\beta_{f}(\omega)+\frac{\omega}{c}n_{2}|\vec{E}|^{2}
+\frac{\omega}{c}\frac{n_{2}\gamma_{xx}}{2}\frac{\partial^{2}}{\partial x^2}(|\vec{E}|^{2})\\
&+\frac{\omega}{c}\frac{n_{2}\gamma_{yy}}{2}\frac{\partial^{2}}{\partial y^2}(|\vec{E}|^{2})+i\alpha_f+i\alpha_f\frac{n_{2}}{n_{0}}|\vec{E}|^{2}\\
&+i\alpha_f\frac{n_{2}}{n_{0}}\frac{\gamma_{xx}}{2}\frac{\partial^{2}}{\partial x^2}(|\vec{E}|^{2})+i\alpha_f\frac{n_{2}}{n_{0}}\frac{\gamma_{yy}}{2}\frac{\partial^{2}}{\partial y^2}(|\vec{E}|^{2})\\
&+\frac{1}{2}\frac{\omega}{c}\frac{\chi_{a}}{n_{0}}.
\end{split}
\end{equation}	
In the framework of Maxwell's equations, we analyze the propagation of optical fields, and  we get the following wave equation for each component of the field vector $\vec{E}$
\begin{equation}\label{eq13}
\Delta \vec{E}-\mu_{0}\frac{\partial^{2}\vec{D}}{\partial t^{2}}=0.
\end{equation}
The operator $\Delta$ is the Laplacian operator whose expression in cartesian coordinates is $\Delta = \frac{\partial^{2}}{\partial x^{2}}+\frac{\partial^{2}}{\partial y^{2}}+\frac{\partial^{2}}{\partial z^{2}}$. The Fourier transform of the quantity $\vec{D}$,  known as the electric displacement vector, is related to that of the electric field via the constitutive relation
\begin{equation}\label{eq14}
\tilde{D}(r,\omega-\omega_{0}) = \epsilon_{ 0} \epsilon( \omega)\tilde{E}(r,\omega-\omega_{0}),
\end{equation}
where $ \epsilon_{ 0}  $ is the vacuum permittivity and $ \tilde{E} $ denotes the Fourier transform of the electric field $\vec{E}$ defined as
\begin{equation}\label{eq15}
\tilde{E}(r,\omega-\omega_{0})=\int_{-\infty}^{\infty} \vec{E}(r,t)\exp(i(\omega-\omega_{0})t)dt.
\end{equation}
In the frequency domain, Eq. (\ref{eq13}) takes the form of the Helmholtz equation
\begin{equation}\label{eq16}
\Delta\tilde{E} + \beta^{2}(\omega)\tilde{E}=0.
\end{equation}
The electric field vector $ \vec{E}(r,t) $ is written as
\begin{equation}\label{eq17}
\vec{E}(r,t)=\frac{1}{2}( \vec{e} \phi(x,y,z,t)
\exp[i(\beta_{0}z-\omega_{0}t)]+c.c.).
\end{equation}
Here, $\phi(x,y,z,t)$ is the slowly varying envelope function which represents the light pulse carrying the information, $c.c.$ is the complex conjugate, $\beta_{0}=n_{0}(\omega_{ 0})k_{0}$ is the propagation constant at the carrier frequency $\omega_{0}$, and $\vec{e}$ is the polarization unit vector. The optical fiber has a cylinder shape with axis $(\emph{oz})$, therefore, the radius $\emph{r}$ is defined as $r= \sqrt{x^2+y^2}$.  For optical beams, the paraxial and quasi-monochromatic approximations correspond to neglecting the $\frac{\partial^{2}}{\partial z^{2}}$ derivatives of the slowly varying amplitude on the grounds that $\left|\frac{\partial^{2} \phi}{\partial z^{2}}\right|\ll\beta_{0}\left|\frac{\partial \phi }{\partial z}\right|$.  This procedure leads to
\begin{equation}\label{eq18}
i\frac{\partial\phi}{\partial z}+\frac{1}{2\beta_{0}}\left( \frac{\partial^{2}\phi}{\partial x^{2}}+\frac{\partial^{2}\phi}{\partial y^2}\right) +(\beta(\omega) 
-\beta_{0}(\omega))\phi=0.
\end{equation}
Fiber dispersion plays a critical role in the propagation of optical pulses because different spectral components associated with pulse broadening can be detrimental to optical communication systems. Mathematically, the effects due to fiber dispersion are accounted for by expanding the mode-propagation coefficient $\beta_{f}(\omega)$ in a Taylor series about the carrier frequency  $\omega_{0}$ at which the pulse spectrum is centred~\cite{FewoJPSJ2008}
\begin{equation}\label{eq19}
\begin{split}
\beta_{f}(\omega,|\phi|^{2})&= \beta_{0}+i\beta_{1}\frac{\partial}{\partial t}-\frac{\beta_{2}}{2}\frac{\partial^{2}}{\partial t^{2}}\\
&+\left( \frac{\partial\beta_{f}}{\partial(|\phi|^{2})}\right) _{0}|\phi|^{2},
\end{split}
\end{equation}
where $(...)_{0}$ denotes the evaluation at $\omega=\omega_{0}$,   $|\phi|^{2}$ = 0,   $\beta_1 = (\frac{\partial\beta}{\partial\omega})_{\omega=\omega_{0}}$ and  $\beta_2  = (\frac{\partial^{2}\beta}{\partial\omega^{2}})_{\omega=\omega_{0}}$.  Hence, the propagation constant $\beta(\omega)$ given in Eq. (\ref{eq12}) becomes
\begin{equation}\label{eq20}
\begin{split}
\beta(\omega)&=\beta_{0}+i\beta_{1}\frac{\partial}{\partial t}-\frac{\beta_{2}}{2}\frac{\partial^{2}}{\partial t^{2}}
+\left( \frac{\partial\beta_{f}}{\partial(|\phi|^{2})}\right) _{0}(|\phi|^{2})\\
&+\frac{\omega}{c}n_{2}|\phi|^{2}+\frac{\omega}{c}\frac{n_{2}\gamma_{xx}}{2}\frac{\partial^{2}}{\partial x^2}(|\phi|^{2})+\frac{\omega}{c}\frac{n_{2}\gamma_{yy}}{2}\frac{\partial^{2}}{\partial y^2}(|\phi|^{2})\\
&+i\alpha_f+i\alpha_f\frac{n_{2}}{n_{0}}|\phi|^{2}+i\alpha_f\frac{n_{2}}{n_{0}}\frac{\gamma_{xx}}{2}\frac{\partial^{2}}{\partial x^2}(|\phi|^{2})\\
&+i\alpha_f\frac{n_{2}}{n_{0}}\frac{\gamma_{yy}}{2}\frac{\partial^{2}}{\partial y^2}(|\phi|^{2})+\frac{1}{2}\frac{\omega}{c}\frac{\chi_{a}}{n_{0}}.
\end{split}
\end{equation}	
Using Eq. (\ref{eq20}) and  substituting $\chi_a$  defined in Eq. (\ref{eq4}), we derive  the following (3+1)D  nonlocal cubic CGL equation:
\begin{equation}\label{eq21}
\begin{split}
i\frac{\partial\phi}{\partial z}&+(\beta_{reff}+i\beta_{ieff})\frac{\partial\phi}{\partial t}+\frac{1}{2\beta_{0}}\left( \frac{\partial^{2}\phi}{\partial x^{2}}+\frac{\partial^{2}\phi}{\partial y^{2}}\right) \\
&+(p_{r}+ip_{i}) \frac{\partial^{2}\phi}{\partial t^{2}}+(\gamma_{r}+i\gamma_{i})
\phi +(q_{r}+iq_{i})|\phi|^{2}\phi\\
&+\left(\gamma_{xx,r}+ i\gamma_{xx,i}\right)\frac{\partial^{2}}{\partial x^2}(|\phi|^{2})\phi\\
&+\left(\gamma_{yy,r}+ i\gamma_{yy,i}\right)\frac{\partial^{2}}{\partial y^2}(|\phi|^{2})\phi=0.
\end{split}
\end{equation}
In the above, $\frac{\partial\phi}{\partial z}$ represents the displacement of the light pulse over the propagation distance $\emph{z}$. The expression  $\left( \frac{\partial^{2}\phi}{\partial x^{2}}+\frac{\partial^{2}\phi}{\partial y^{2}}\right) $ corresponds to the linear diffraction of the light pulse along the transverse directions $\emph{x}$ and $\emph{y}$. The terms $ \frac{\partial^{2}}{\partial x^2}(|\phi|^{2})\phi $  and  $ \frac{\partial^{2}}{\partial y^2}(|\phi|^{2})\phi $ correspond to the nonlinear diffraction related to spatial nonlocality. This nonlinear diffraction is associated with linear diffraction in order to influence self-focusing and avoid collapse~\cite{BezuhanovPRA2008}. The parameters $\beta_{reff}$ = $-\frac{g_{p}}{2n_{0}}\frac{2\delta T_{2}}{(1+\delta^2)^{2}}$ and $\beta_{ieff}$ = $\beta_{1}+\frac{g_{p}}{2n_{0}}\frac{(1-\delta^{2})T_{2}}{(1+\delta^2)^{2}}$, are the real and imaginary parts of the inverse of the group velocity, respectively. The coefficient $p_{r}= -\frac{\beta_{2}}{2}-\frac{g_{p}}{2n_{0}}\frac{\delta(\delta^{2}-3)T_{2}^{2}}{(1+\delta^{2})^{3}}$, measures the wave dispersion, and           $p_{i}=-\frac{g_{p}}{2n_{0}}\frac{(1-3\delta^{2})T_{2}^{2}}{(1+\delta^{2})^{3}}$ is the spectral filtering.  The term $\gamma_{r}=\frac{g_{p}}{2n_{0}}\frac{\delta}{(1+\delta^{2})}$ is the  linear loss/gain, and  $\gamma_{i}=\alpha_{f}-\frac{g_{p}}{2n_{0}}\frac{1}{(1+\delta^{2})}$ is  the  frequency shift. The parameter $q_{r}=n_{2}\frac{\omega }{c}+\left( \frac{\partial\beta_{f}}{\partial(|\phi|^{2})}\right) _{0}$ represents the nonlinear coefficient. The case $q_{r}$$>$$ 0$ corresponds to the self-focusing Kerr nonlinearity, while the case $q_{r}$$<$$ 0$ corresponds to the self-defocusing Kerr nonlinearity. The parameter $q_{i}=n_{2}\frac{\alpha_{f} }{n_{0}}$, accounts for nonlinear gain (loss) and/or other amplification (absorption) processes. Expressions $\gamma_{xx,r}=\frac{1}{2}\frac{n_{2} \omega}{c}\gamma_{xx}$,  $\gamma_{yy,r}=\frac{1}{2}\frac{n_{2} \omega}{c}\gamma_{yy}$,  $\gamma_{xx,i}=\frac{1}{2}\frac{n_{2} \alpha_{f} }{n_{0}} \gamma_{xx} $, and  $\gamma_{yy,i}=\frac{1}{2}\frac{n_{2} \alpha_{f} }{n_{0}} \gamma_{yy} $, represent the real and imaginary parts of the nonlocality degrees along the transverse coordinates $x$ and $y$, respectively. 

To scale Eq. (\ref{eq21}), we introduce the following physical parameters, namely, the diffraction length or Rayleigh length in the homogeneous medium: $L_{Diff}=\beta_{0}r_{0}^{2}$, where  $r_{0}$ is the beam radius. The dispersion length  $L_{Disp}=\frac{T_{0}^{2}}{|p_{r}|}$, with  $T_{0}$ representing the typical initial pulse width. The effective length characterizing the influence of the nonlinearity $L_{NL}=\frac{1}{q_{r} P_{0}}$, where $P_{{0}}$ is the peak power of the incident pulse. The transverse coordinates scale are $X=\frac{x}{r_{0}}$	and $Y=\frac{y}{r_{0}}$, respectively. The longitudinal coordinate scale  $Z=\frac{z}{L_{Disp}}$, and the temporal coordinate scale  $\tau=\frac{T}{T_{0}}$. Here, $T=t-\frac{z}{v_g}$ is the time in the moving coordinate system. The normalized field amplitude is $\Phi(X,Y,Z,\tau)=\sqrt{P_{0}} N \phi(x,y,z,t)$, where $N^{2}=\frac{L_{Disp}}{L_{NL}}$, with $N=\sqrt{\frac{q_{r}P_{0}T_{0}^{2}}{|p_{r}|}}$. We also set $a_{1}=\frac{p_{i}}{p_{r}}$, $b_{1}=\frac{\gamma_{i}}{\gamma_{r}}$, $c_{0}=\frac{q_{i}}{q_{r}}$, $c_{_{_{1,XX}}}=\frac{\gamma_{_{_{XX,i}}}}{\gamma_{_{_{XX,r}}}}$, and $c_{_{_{2,YY}}}=\frac{\gamma_{_{_{YY,i}}}}{\gamma_{_{_{YY,r}}}}$.
   	
Taking into account these scaling transformations, Eq.~(\ref{eq21}) then takes the form
\begin{equation}\label{eq22}
\begin{split}
i\frac{\partial\Phi}{\partial Z}&+\zeta_{1}\left( \frac{\partial^{2}\Phi}{\partial X^{2}}+\frac{\partial^{2}\Phi}{\partial Y^{2}}\right) +\left(1+ ia_{1}\right)\zeta_{2} \frac{\partial^{2}\Phi}{\partial \tau^{2}}\\
&+\left(1+ ib_{1}\right)\zeta_{3}\Phi +\left(1+ ic_{0}\right)\zeta_{4}|\Phi|^{2}\Phi\\
&+\left(1+ic_{_{_{1,XX}}}\right)\zeta_{5}\frac{\partial^{2}}{\partial X^2}(|\Phi|^{2})\Phi\\
&+\left(1+ ic_{_{_{2,YY}}}\right)\zeta_{6}\frac{\partial^{2}}{\partial Y^2}(|\Phi|^{2})\Phi=0,
\end{split}
\end{equation}
where $\zeta_{1}=\frac{L_{Disp}}{\beta_{0} r_{0}^{2}}$, $\zeta_{2}=\frac{p_{r} L_{Disp}}{T_{0}^{2}N\sqrt{P_{0}} }$, $\zeta_{3}=\gamma_{r}L_{Disp}$, $\zeta_{4}=\frac{q_{r}L_{NL}}{P_{0}}$, $\zeta_{5}=\frac{\gamma_{_{_{XX}}}L_{NL}}{P_{0} r_{0}^{2}}$, and $\zeta_{6}=\frac{\gamma_{_{_{YY}}}L_{NL}}{P_{0} r_{0}^{2}}$. Considering Eq. (24), for  $a_{1}$ = $b_{1}$ = $c_{0}$ = $c_{_{_{1,XX}}}$ = $c_{_{_{2,YY}}}$	= $0$, and neglecting the second-order dispersion term $\frac{\partial^{2}\Phi}{\partial \tau^{2}}$, we recover the (2+1)D nonlocal nonlinear Schr\"{o}dinger (NLS) equation that was derived by Bezuhanov et al.~\cite{BezuhanovPRA2008}  in the limit of weak nonlocality. The conditions for breathing soliton formation in one and two transverse dimensions were established for this equation. Furthermore, it was shown that the interplay between nonlinear diffraction and self-focusing is found to result in an increase of the power needed to form nonlocal spatial solitons. Indeed, the nonlocal 2D NLS equation has been also derived by Skupin et al.~\cite{SkupinPRE2006}  in the highly nonlocal limit. Due to the mixture of local and nonlocal types of nonlinearity (Gaussian model of nonlocality, thermal nonlinearity, and the model of a dipolar Bose-Einstein condensate), a variety of solutions, such as rotating and nonrotating azimuthons, accessible solitons can be stabilized. Moreover, the stabilization of 2D ring dark solitons and ring anti-dark solitons was demonstrated in nonlocal media~\cite{ChenJPB2017}.
		
\section{Analytical treatment using variational approach}\label{sec3}
		
In this section, we employ the variational method~\cite{KaupPhysD1995,AnkiewiczOFTech2007,SkarkaPRL2010} for dissipative systems to look for approximated solutions to Eq.~(\ref{eq22}) and obtain physical insight in terms of a few relevant parameters that will be then implemented in numerical simulations to confirm the analytic predictions qualitatively. In order to describe the dynamics of pulse evolution, various treatments have been developed to extract approximated soliton solutions to integrable and non-integrable nonlinear partial differential equations and have received many different names depending on the field of application, namely the method of moments~\cite{DindaPRE2001}, the method of collective coordinates~\cite{FewoJPSJ2008,EssamaPRE2014,KamagatePRE2009}, the time-dependent variational method~\cite{AnkiewiczOFTech2007}, the effective-particle method~\cite{MontesinosChaos2005,MorseBook1953}, averaged Lagrangian description~\cite{SkarkaPRL2010,SkarkaPRL2006}, to name a few. Lagrangian methods have become widely accepted as the preferred approach to account for the dynamics of the light pulse in optical fiber. Thus, in order to use the variational approach, the (3+1)D nonlocal cubic CGL equation can be rewritten in the form
\begin{equation}\label{eq23}
\begin{split}
i\frac{\partial\Phi}{\partial Z}&+\zeta_{1}\left( \frac{\partial^{2}\Phi}{\partial X^{2}}+\frac{\partial^{2}\Phi}{\partial Y^{2}}\right) +\zeta_{2} \frac{\partial^{2}\Phi}{\partial \tau^{2}}+\zeta_{3}\Phi\\
& +\zeta_{4}|\Phi|^{2}\Phi+\zeta_{5}\frac{\partial^{2}}{\partial X^2}(|\Phi|^{2})\Phi\\
&+\zeta_{6}\frac{\partial^{2}}{\partial Y^2}(|\Phi|^{2})\Phi=\mathcal{Q},
\end{split}
\end{equation}	
in which the right-hand side contains dissipative terms
\begin{equation}\label{eq24}
\begin{split}
\mathcal{Q}&=- ia_{1}\zeta_{2} \frac{\partial^{2}\Phi}{\partial \tau^{2}}-ib_{1}\zeta_{3}\Phi -ic_{0}\zeta_{4}|\Phi|^{2}\Phi\\		
&-ic_{_{_{1,XX}}}\zeta_{5}\frac{\partial^{2}}{\partial X^2}(|\Phi|^{2})\Phi-ic_{_{_{2,YY}}}\zeta_{6}\frac{\partial^{2}}{\partial Y^2}(|\Phi|^{2})\Phi.
\end{split}
\end{equation}	
It should be noted that defining an appropriate ansatz function is essential for using the variational approach. In doing so, let us consider the trial function of the Gaussian shape
\begin{equation}\label{eq25}
\begin{split}
\Phi(X,Y,Z,\tau)&=A(Z)\exp\Big(-\frac{X^{2}}{\sigma_{_X}^{2}(Z)}-\frac{Y^{2}}{\sigma_{_Y}^{2}(Z)}\\
&-\frac{\tau^{2}}{\sigma_{\tau}^{2}(Z)}+\frac{ik_{0}}{2}(\vartheta_{_{X}}(Z)X^{2}\\
&+\vartheta_{_{Y}}(Z)Y^{2}+\vartheta_{\tau}(Z)\tau^{2})+i\psi(Z)\Big),
\end{split}
\end{equation}
where  $ A(Z)$ is the amplitude, $\sigma_{_X}(Z)$ and $\sigma_{_Y}(Z)$ are the beamwidths  in the transverse coordinates $(\emph{X},\emph{Y})$, $\sigma_{\tau}(Z)$ is the temporal beamwidth, $\vartheta_{_{X}}(Z)$ and $\vartheta_{_{Y}}(Z)$ are the wave-front curvatures along the transverse coordinates $(\emph{X},\emph{Y})$, $\vartheta_{\tau}(Z)$ is the temporal wave-front curvature, $\psi(Z)$ is the phase, and $k_{0}$ is the wavenumber. Variables $A(Z)$, $\sigma_{_X}(Z)$, $\sigma_{_Y}(Z)$, $\sigma_{\tau}(Z)$, $\vartheta_{_{X}}(Z)$, $\vartheta_{_{Y}}(Z)$, $\vartheta_{\tau}(Z)$ and $\psi(Z)$ are all the parameters of the light pulse. Then,  we obtain the first-order differential equations (FODEs) of these parameters, which describe the evolution of the light pulse in the optical fiber. The  variational approach used in this study is based on the Euler-Lagrange equation~\cite{SkarkaPiers2007,GarciaPRE2000}
\begin{equation}\label{eq26}
\begin{split}
\frac{d}{dZ}\left( \frac{\partial \langle L_{c}\rangle}{\partial q'}\right) &-\frac{\partial \langle L_{c}\rangle}{\partial q}\\
&=2{\rm Re}\left\{\int\int\int dXdYd\tau\mathcal{Q}\frac{\partial\Phi^{\ast}}{\partial q}\right\},
\end{split}
\end{equation}
where $\Phi^{\ast}$ is the complex conjugate of the ansatz function $\Phi$,  $\rm Re\{\cdot\}$ denotes the real part, and $\emph{q}$ is the variable that corresponds to all the parameters of the light pulse ($A(Z)$, $\sigma_{_X}(Z)$, $\sigma_{_Y}(Z)$, $\sigma_{\tau}(Z)$, $\vartheta_{_{X}}(Z)$, $\vartheta_{_{Y}}(Z)$, $\vartheta_{\tau}(Z)$ , $\psi(Z)$), with $(q'=\frac{dq}{dZ})$. The left-hand side of Eq.~(\ref{eq23}), represents the conservative Lagrange's equation, with the conservative Lagrangian $L_{c}$ being given by
\begin{equation}\label{eq27}
\begin{split}
{L}_{c}&= i\frac{\partial\Phi}{\partial Z}+\zeta_{1}\left( \frac{\partial^{2}\Phi}{\partial X^{2}}+\frac{\partial^{2}\Phi}{\partial Y^{2}}\right) +\zeta_{2} \frac{\partial^{2}\Phi}{\partial \tau^{2}}\\
&+\zeta_{3}\Phi +\zeta_{4}|\Phi|^{2}\Phi+\zeta_{5}\frac{\partial^{2}}{\partial X^2}(|\Phi|^{2})\Phi\\
&+\zeta_{6}\frac{\partial^{2}}{\partial Y^2}(|\Phi|^{2})\Phi. 
\end{split}
\end{equation}
Moreover, $\langle \textit{L}_{c}\rangle$  is the Lagrangian density written as 
\begin{align}\label{eq28}
\langle L_{c}\rangle=\int\int\int L_{c} dX dY d\tau,
\end{align}
and the right-hand side $\mathcal{Q}$ of Eq.(\ref{eq23}) is the dissipative term. Using Eqs.~(\ref{eq24})-(\ref{eq28}),  we get the following set of coupled first-order differential equations resulting from the variation with respect to the light pulse parameters:
\begin{subequations}\label{eq29}
\begin{equation}\label{eq29:a}
\begin{split}
\frac{dA}{dZ}&=-A\zeta_{1}k_{0}\vartheta_{_{X}}-A\zeta_{1}k_{0}\vartheta_{_{Y}}-A\zeta_{2}\vartheta_{\tau}
-\frac{7}{2}Ab_{1}\zeta_{3}\\
&-\frac{77}{16}A^3\sqrt{2}c_{0}\zeta_{4}+\frac{1}{8}Aa_{1}\zeta_{2}\left(35k_{0}^2\vartheta_{\tau}^2\sigma_{\tau}^2+\frac{156}{\sigma^2}\right)\\
&+\frac{1}{128}A^3\sqrt{2}c_{_{1,XX}}\zeta_{5}\left(75k_{0}^2\vartheta_{_{X}}^2\sigma_{_X}^2+\frac{996}{\sigma_{_X}^2}\right)\\
&+\frac{1}{128}A^3\sqrt{2}c_{_{2,YY}}\zeta_{6}\left(75k_{0}^2\vartheta_{_{Y}}^2\sigma_{_Y}^2+\frac{996}{\sigma_{_Y}^2}\right),
\end{split}
\end{equation}	
\begin{equation}\label{eq29:b}
\begin{split}
\frac{d\sigma_{_X}}{dZ}&=2\zeta_{1}k_{0}\vartheta_{_{X}}\sigma_{_X}+b_{1}\zeta_{3}\sigma_{_X}+\frac{15}{8}A^2\sqrt{2}c_{0}\zeta_{4}\sigma_{_X}\\
&+\frac{1}{4}a_{1}\zeta_{2}\left(-7k_{0}^2\vartheta_{\tau}^2\sigma_{\tau}^2-\frac{28}{\sigma_{\tau}^2}\right)\sigma_{_X}\\
&+\frac{1}{64}A^2\sqrt{2}c_{_{1,XX}}\zeta_{5}\left(13k_{0}^2\vartheta_{_{X}}^2\sigma_{_X}^2-\frac{252}{\sigma_{_X}^2}\right)\sigma_{_X}\\
&+\frac{15}{64}A^2\sqrt{2}c_{_{2,YY}}\zeta_{6}\left(-k_{0}^2\vartheta_{_{Y}}^2\sigma_{_Y}^2-\frac{12}{\sigma_{_Y}^2}\right)\sigma_{_X},
\end{split}
\end{equation}
\begin{equation}\label{eq29:c}
\begin{split}
\frac{d\sigma_{_Y}}{dZ}&=2\zeta_{1}k_{0}\vartheta_{_{Y}}\sigma_{_Y}+b_{1}\zeta_{3}\sigma_{_Y}+\frac{15}{8}A^2\sqrt{2}c_{0}\zeta_{4}\sigma_{_Y}\\
&+\frac{1}{4}a_{1}\zeta_{2}\left(-7k_{0}^2\vartheta_{\tau}^2\sigma_{\tau}^2-\frac{28}{\sigma_{\tau}^2}\right)\sigma_{_Y}\\
&+\frac{15}{64}A^2\sqrt{2}c_{_{1,XX}}\zeta_{5}\left(-k_{0}^2\vartheta_{_{X}}^2\sigma_{_X}^2-\frac{12}{\sigma_{_X}^2}\right)\sigma_{_Y}\\
&+\frac{1}{64}A^2\sqrt{2}c_{_{2,YY}}\zeta_{6}\left(13k_{0}^2\vartheta_{_{Y}}^2\sigma_{_Y}^2-\frac{252}{\sigma_{_Y}^2}\right)\sigma_{_Y},
\end{split}
\end{equation}
\begin{equation}\label{eq29:d}
\begin{split}
\frac{d\sigma_{\tau}}{dZ}&=2\zeta_{2}\vartheta_{\tau}\sigma_{\tau}+b_{1}\zeta_{3}\sigma_{\tau}+\frac{15}{8}A^2\sqrt{2}c_{0}\zeta_{4}\sigma_{\tau}\\
&+\frac{1}{4}a_{1}\zeta_{2}\left(-5k_{0}^2\vartheta_{\tau}^2\sigma_{\tau}^2-\frac{36}{\sigma_{\tau}^2}\right)\sigma_{\tau}\\
&+\frac{15}{64}A^2\sqrt{2}c_{_{1,XX}}\zeta_{5}\left(-k_{0}^2\vartheta_{_{X}}^2\sigma_{_X}^2-\frac{12}{\sigma_{_X}^2}\right)\sigma_{\tau}\\
&+\frac{15}{64}A^2\sqrt{2}c_{_{2,YY}}\zeta_{6}\left(-k_{0}^2\vartheta_{_{Y}}^2\sigma_{_Y}^2-\frac{12}{\sigma_{_Y}^2}\right)\sigma_{\tau},
\end{split}
\end{equation}
\begin{equation}\label{eq29:e}
\begin{split}
\frac{d\vartheta_{_{X}}}{dZ}&=A^2\sqrt{2}\zeta_{4}\frac{1}{k_{0}\sigma_{_X}^2}+8\zeta_{1}\frac{1}{\sigma_{_X}^4}-2\zeta_{1}k_{0}\vartheta_{_{X}}^2\\
&+\frac{1}{36}\frac{A^2\sqrt{2}c_{_{1,XX}}\zeta_{5}}{k_{0}\sigma_{_X}^2}\left(\frac{216}{\sigma_{_X}^2}+127k_{0}\vartheta_{_{X}}\right)\\
&+\frac{1}{36}\frac{A^2\sqrt{2}c_{_{2,YY}}\zeta_{6}}{k_{0}\sigma_{_X}^2}\left(\frac{72}{\sigma_{_Y}^2}+73k_{0}\vartheta_{_{Y}}\right),
\end{split}
\end{equation}
\begin{equation}\label{eq29:f}
\begin{split}
\frac{d\vartheta_{_{Y}}}{dZ}&=A^2\sqrt{2}\zeta_{4}\frac{1}{k_{0}\sigma_{_Y}^2}+8\zeta_{1}\frac{1}{\sigma_{_Y}^4}-2\zeta_{1}k_{0}\vartheta_{_{Y}}^2\\
&+\frac{1}{36}\frac{A^2\sqrt{2}c_{_{1,XX}}\zeta_{5}}{k_{0}\sigma_{_Y}^2}\left(\frac{72}{\sigma_{_X}^2}+73k_{0}\vartheta_{_{X}}\right)\\
&+\frac{1}{36}\frac{A^2\sqrt{2}c_{_{2,YY}}\zeta_{6}}{k_{0}\sigma_{_Y}^2}\left(\frac{216}{\sigma_{_Y}^2}+127k_{0}\vartheta_{_{Y}}\right),
\end{split}
\end{equation}
\begin{equation}\label{eq29:g}
\begin{split}
\frac{d\vartheta_{\tau}}{dZ}&=A^2\sqrt{2}\zeta_{4}\frac{1}{k_{0}\sigma_{\tau}^2}+8a_{1}\zeta_{2}\frac{\vartheta_{\tau}}{\sigma_{\tau}^2}\\
&-8\zeta_{2}\frac{1}{k_{0}\sigma_{\tau}^4}-2\zeta_{2}k_{0}\vartheta_{\tau}^2\\
&+\frac{1}{36}\frac{A^2\sqrt{2}c_{_{1,XX}}\zeta_{5}}{k_{0}\sigma_{\tau}^2}\left(\frac{72}{\sigma_{_X}^2}+73k_{0}\vartheta_{_{X}}\right)\\
&+\frac{1}{36}\frac{A^2\sqrt{2}c_{_{2,YY}}\zeta_{6}}{k_{0}\sigma_{\tau}^2}\left(\frac{72}{\sigma_{_Y}^2}+73k_{0}\vartheta_{_{Y}}\right),
\end{split}
\end{equation}
\begin{equation}\label{eq29:h}
\begin{split}
\frac{d\psi}{dZ}&=-2\frac{\zeta_{1}}{\sigma_{_X}^2}-2\frac{\zeta_{1}}{\sigma_{_Y}^2}-2\frac{\zeta_{2}}{\sigma_{\tau}^2}\\
&-\frac{7}{8}A^2\sqrt{2}\zeta_{4}-\zeta_{3}-a_{1}\zeta_{2}k_{0}\vartheta_{\tau}\\
&+\frac{1}{288}A^2\sqrt{2}c_{_{1,XX}}\zeta_{5}\left(-\frac{648}{\sigma_{_X}^2}-401k_{0}\vartheta_{_{X}}\right)\\
&+\frac{1}{288}A^2\sqrt{2}c_{_{2,YY}}\zeta_{6}\left(-\frac{648}{\sigma_{_Y}^2}-401k_{0}\vartheta_{_{Y}}\right).
\end{split}
\end{equation}
\end{subequations}
Eq. (\ref{eq29:a})-(\ref{eq29:h}) constitute a complete set of variable parameters characterizing dissipative light bullets in optical fiber amplifiers under the suitable competition between dopants and a spatially nonlocal nonlinear response. To reach the steady-state solutions of the system of Eqs.(\ref{eq29:a})-(\ref{eq29:g}), we set $Z$ derivatives of the light pulse parameters to zero. The problem of light pulse instabilities that depend on the number of space dimensions and nonlinearity strength has recently attracted considerable attention. In fact, the existence and stability of multidimensional optical dissipative soliton solutions of the cubic-quintic CGL equation were addressed and comprehensively analyzed~\cite{SkarkaPRL2010}. In the dissipative case, it was already remarked that the family of solutions reduces to a fixed double solution for a given set of dissipative parameters~\cite{SkarkaJOA2008}. Indeed, only symmetric steady-state solutions with equal spatial widths,  curvatures, and nonlocality degrees can exist, which means $\sigma_{_X}=\sigma_{_{Y}}=\sigma_{_{X=Y}}=\sigma$, $\vartheta_{_{X}}=\vartheta_{_{Y}}=\vartheta_{_{X=Y}}=\vartheta$, $\gamma_{_{XX}}=\gamma_{_{YY}}=\gamma_{_{X=Y}}$, $\zeta_{5}=\zeta_{6}=\zeta_{_{X=Y}}$ and  $c_{_{1,XX}}=c_{_{2,YY}}=c_{_{X=Y}}$. IUnder such conditions, the amplitude as a steady state solution of Eqs.(\ref{eq29:a})-(\ref{eq29:g}) has two discrete values  $A_{+}$ and $A_{-}$ given by
\begin{equation}\label{eq30}
A_{\pm}=\sqrt{\frac{-D\pm\sqrt{D^{2}-4BC}}{2B}}+O(\theta^{2}),
\end{equation}
where
\begin{equation}\label{eq31}
\begin{split}
B&=\sqrt{2}\zeta_{4}\zeta_{_{X=Y}}c_{_{X=Y}}(240a_{1}-496c_{0}-249),\\
D&=-320b_{1}\zeta_{3}\zeta_{_{X=Y}}c_{_{X=Y}}+\zeta_{1}\zeta_{4}(128a_{1}-256c_{0}),\\
C&=-128\sqrt{2}b_{1}\zeta_{1}\zeta_{3},
\end{split}
\end{equation}
and $\theta$ = $\text{max}$ $\left( |a_{1}\zeta_{2}|  ,  |b_{1}\zeta_{3}|,  |c_{0}\zeta_{4}|, |c_{{X=Y}}\zeta_{_{X=Y}}| \right)$. The other relevant stationary solutions are explicitly given by the following four-parameter family of the dissipative light bullets:\\
{\it (i) The beamwidth in the transverse coordinates}
\begin{equation}\label{eq32}
\sigma_{_{X=Y}}=\frac{2\sqrt{-\zeta_{4}\left(2\zeta_{_{X=Y}}c_{_{X=Y}}A^2+\sqrt{2}\zeta_{1}\right)}}{A\zeta_{4}}+O(\theta^{2}),
\end{equation}
{\it (ii) The temporal beamwidth}
\begin{equation}\label{eq33}
\sigma_{\tau}=\frac{2\sqrt{\zeta_{4}\sqrt{2}\zeta_{2}}}{A\zeta_{4}}+O(\theta^{2}),
\end{equation}
{\it (iii) The wave-front curvatures along the transverse coordinates}
\begin{equation}\label{eq34}
\begin{split}
\vartheta_{_{X=Y}}&=\frac{1}{32 k_{0}\zeta_{1}\left(2A^2\zeta_{_{X=Y}}c_{_{X=Y}}+\sqrt{2}\zeta_{1}\right)}\\
&\times\left(28A^2a_{1}\zeta_{1}\zeta_{4}-60A^2c_{0}\zeta_{1}\zeta_{4}\right.\\&\left.-32A^2b_{1}\zeta_{3}\zeta_{_{X=Y}}c_{_{X=Y}}-16\sqrt{2}b_{1}\zeta_{1}\zeta_{3}\right)+O(\theta^{2}),
\end{split}
\end{equation}
{\it (iv) The temporal wave-front curvature}
\begin{equation}\label{eq35}
\vartheta_{\tau}=\frac{\left(9A^2\sqrt{2}a_{1}\zeta_{4}-15A^2\sqrt{2}c_{0}\zeta_{4}-8b_{1}\zeta_{3}\right)}{16\zeta_{2}}+O(\theta^{2}).
\end{equation}
To further proceed, we first need to make a remark about the conservative systems. In this case, the total energy or beam power is conserved. For dissipative systems, the total energy or the beam power $  P=(\frac{\pi}{2})^{3/2}A^{2}\sigma_{_{X=Y}}^2\sigma_{\tau}$  is no longer conserved contrary to the case of conservative systems, but evolves in the so-called balance equation with nonzero curvatures. Therefore, using steady-state solutions given in Eqs.~(\ref{eq32}) and (\ref{eq33}), the total power corresponding to ansatz (\ref{eq25}) for the non-conservative (3+1)D nonlocal cubic CGL equation (\ref{eq23}) is given by~\cite{SkarkaJOA2008}
\begin{equation}\label{eq36}
P=\frac{8\left(2\zeta_{_{X=Y}}c_{_{X=Y}}A^2+\sqrt{2}\zeta_{1}\right)\sqrt{\zeta_{4}\sqrt{2}\zeta_{2}}}{\left(\frac{\pi}{2}\right)^{-3/2}A\zeta_{4}^2}.
\end{equation}
It is also worth mentioning that the simultaneous balance between linear diffraction, dispersion, and nonlinear diffraction is induced by the spatial nonlocality and between gain and loss, depending on fixed steady-state solutions with nonzero spatial and temporal curvatures. 
\begin{figure}[t]
\centering
\includegraphics[width=3.50in]{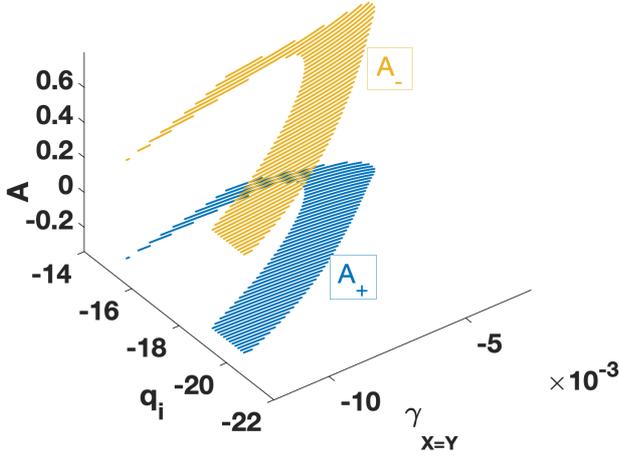}
\caption{Double solutions $A_{+}$  and $A_{-}$ of the steady state amplitudes versus $q_{i}$ and  $\gamma_{_{X=Y}}$. The solution $ A_{-} $ (yellow sheet) is stable. The solution $ A_{+} $  (blue sheet) is rather unstable. The following parameter values have been used: $N_{r}=2^{1/2}(4/3)^{3/4}$ (the normalized coefficient),  $p_{r}=0.579/N_{r}$ (the anomalous dispersion),  $p_{i}=0.201/N_{r}$ (the spectral filtering), $q_{r}=-110/N_{r}$ (the self-defocusing Kerr nonlinearity),  $\gamma_{r}=150/N_{r}$ (the linear gain),  $\gamma_{i}=-530/N_{r}$ (the frequency shift),   and $k_{0}=2\pi/1.55$ (the wavenumber).}
\label{fig1}
\end{figure}
\begin{figure*}[t]
\centering
\includegraphics[width=7.50in]{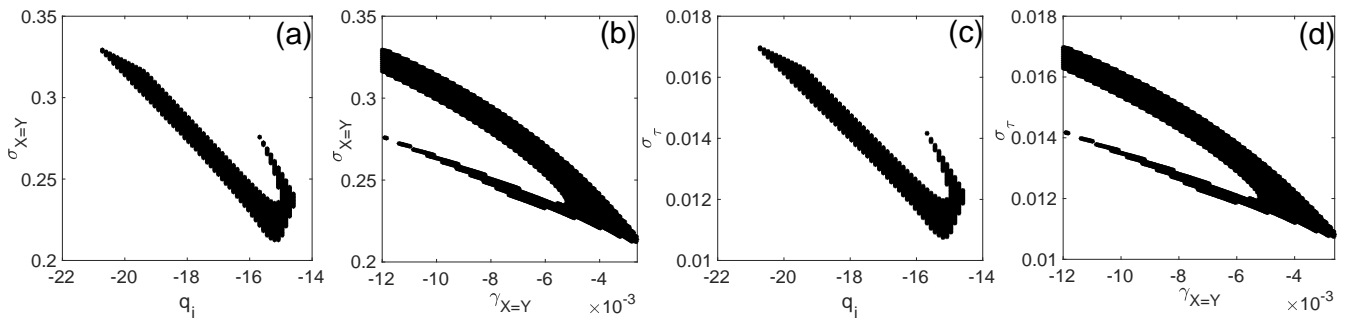}
\caption{Diagrams of stability in the $(q_{i}, \sigma_{_{X=Y}})-$plane [panel (a)], the $(\gamma_{_{X=Y}},\sigma_{_{X=Y}})-$plane [panel (b)], the $(q_{i}, \sigma_{\tau})-$plane [panel(c)] and the  $(\gamma_{_{X=Y}},\sigma_{\tau})-$plane [panel(d)], where the dark zone stands for the stable solution $A_{-}$ under anomalous dispersion. The parameter values used are the same as in Fig.~\ref{fig1}. }
\label{fig2}
\end{figure*}
\begin{figure*}[t]	
\centering
\includegraphics[width=7.50in]{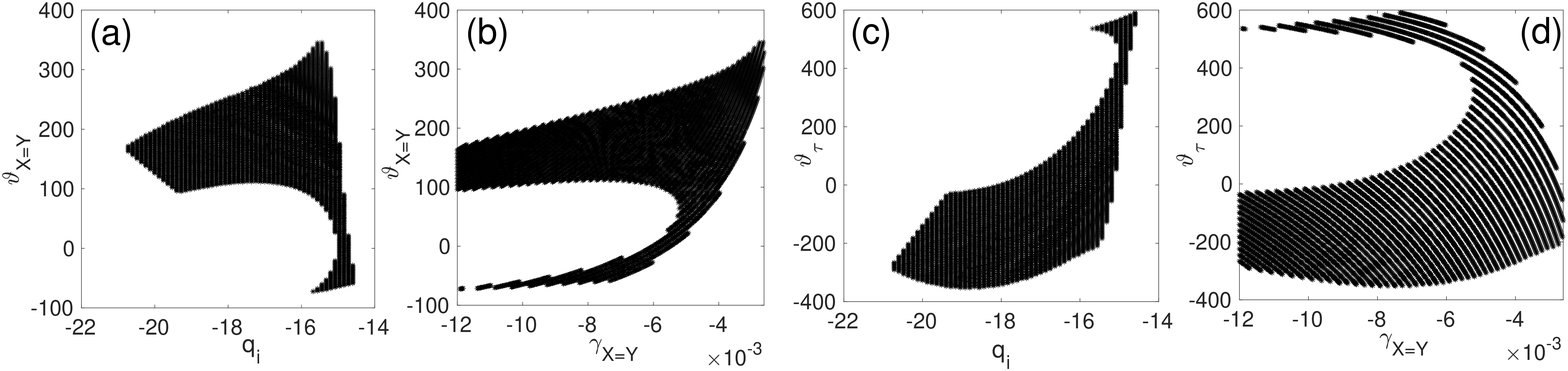}
\caption{Diagrams of stability in the $(q_{i}, \vartheta_{_{X=Y}})-$plane [panel (a)], the $(\gamma_{_{X=Y}},\vartheta_{_{X=Y}})-$plane [panel (b)], the $(q_{i}, \vartheta_{\tau})-$plane [panel(c)] and the  $(\gamma_{_{X=Y}},\vartheta_{\tau})-$plane [panel(d)], where the dark zone stands for the stable solution $A_{-}$ under anomalous dispersion. The parameter values used are the same as in Fig.~\ref{fig1}.}
\label{fig3}
\end{figure*}

\section{Stability criterion for steady-state solutions}\label{sec4}

In order to picture the stability zone, we use the Routh-Hurwitz stability criterion described by a necessary condition and a sufficient condition. Applying this stability criterion requires that we construct a Jacobi determinant. From the latter,  we get the polynomial characteristic equation, and we are then able to verify the necessary and sufficient condition of the Routh-Hurwitz stability criterion. In the following,  we introduce the  notations: $ F_{A}\equiv\frac{dA}{dZ}$, $ F_{\sigma}\equiv \frac{d\sigma}{dZ}$, $ F_{\vartheta}\equiv \frac{d\vartheta}{dZ}$, $ F_{\sigma_{\tau}}\equiv \frac{d\sigma_{\tau}}{dZ}$, and  $ F_{\vartheta_{\tau}} \equiv\frac{d\vartheta_{\tau}}{dZ}$, resulting from  Eqs.~(\ref{eq29:a})-(\ref{eq29:g}). Recall that in the case of a symmetric input, the set of Eqs.~(\ref{eq29:a})-(\ref{eq29:g}) is reduced to only five equations: (\ref{eq29:a}), (\ref{eq29:b}), (\ref{eq29:d}), (\ref{eq29:e}) and (\ref{eq29:g}). The Jacobi determinant is constructed from the derivatives of these terms  $ F_{A}$, $ F_{\sigma}$, $F_{\vartheta}$, $ F_{\sigma_{\tau}}$, and $ F_{\vartheta_{\tau}}$, with respect to amplitude, spatial and temporal widths, spatial and temporal curvatures  taken at the equilibrium state
\begin{multline}\label{eq37}
{\rm det}(J-\lambda I)=\\
\begin{vmatrix}
\frac{\partial F_{A}}{\partial A }-\lambda& \frac{\partial F_{A}}{\partial \sigma } & \frac{\partial F_{A}}{\partial \vartheta} & \frac{\partial F_{A}}{\partial \sigma_{\tau} } & \frac{\partial F_{A}}{\partial \vartheta_{\tau} } \\ 
\frac{\partial F_{\sigma}}{\partial A }&\frac{\partial F_{\sigma}}{\partial \sigma }-\lambda & \frac{\partial F_{\sigma}}{\partial \vartheta } & \frac{\partial F_{\sigma}}{\partial \sigma_{\tau}  } & \frac{\partial F_{\sigma}}{\partial \vartheta_{\tau} }  \\ 
\frac{\partial F_{\vartheta}}{\partial A }& \frac{\partial F_{\vartheta}}{\partial \sigma } & \frac{\partial F_{\vartheta}}{\partial \vartheta }-\lambda & \frac{\partial F_{\vartheta}}{\partial \sigma_{\tau}  }  & \frac{\partial F_{\vartheta}}{\partial \vartheta_{\tau}  } \\ 
\frac{\partial F_{\sigma_{\tau}}}{\partial A }& \frac{\partial F_{\sigma_{\tau}}}{\partial \sigma} &\frac{\partial F_{\sigma_{\tau}}}{\partial \vartheta}  & \frac{\partial F_{\sigma_{\tau}}}{\partial \sigma_{\tau}  }-\lambda & \frac{\partial F_{\sigma_{\tau}}}{\partial \vartheta_{\tau}  }  \\ 
\frac{\partial F_{\vartheta_{\tau}}}{\partial A }& \frac{\partial F_{\vartheta_{\tau}}}{\partial \sigma } & \frac{\partial F_{\vartheta_{\tau}}}{\partial \vartheta } & \frac{\partial F_{\vartheta_{\tau}}}{\partial F_{\sigma_{\tau}}}  & \frac{\partial F_{\vartheta_{\tau}}}{\partial \vartheta_{\tau}  }-\lambda
\end{vmatrix} \\=0,	
\end{multline}
where $I$ is the identity matrix. The fifth-order characteristic polynomial obtained from the Jacobi determinant is given by 
 \begin{equation}\label{eq38}
\lambda^{5}+a_{1}\lambda^{4}+ a_{2}\lambda^{3}+a_{3}\lambda^{2}+a_{4}\lambda+a_{5}=0.
 \end{equation}
 The coefficients $a_{1},...,a_{5}$, depend on the partial derivatives of the functions  $F_{A} $,  $F_{\sigma} $,   $F_{\vartheta} $,  $ F_{\sigma_{\tau}}$, and  $ F_{\vartheta_{\tau}}$  given in the Appendix. However, the analytical expressions of those coefficients could not be presented due to their complicated expressions. For stability to be reached, the necessary condition implies that all the roots of the  characteristic Eq. (\ref{eq38}) should have negative real parts, i.e.,  $ {\rm Re}(\lambda_{i}) $ $ < $ $0$, with $i=1$,...,$5$.  The sufficient condition based on the Routh-Hurwitz criterion is such that the coefficients $a_{1},...,a_{5}$ and their combinations should be positive~\cite{SkarkaPRL2006,DjazetEPJD2020}, i.e.,
\begin{equation}\label{eq39}
a_{i}> 0,\;\; \text{with }\;\; i =1,...,5,
\end{equation}
and
\begin{equation}\label{eq40}
\begin{split}
 b_{1}&=\frac{a_{1}a_{2}-a_{3}}{a_{1}},\;\;
 b_{3}=\frac{a_{1}a_{4}-a_{5}}{a_{1}},\\
 c_{1}&=\frac{b_{1}a_{3}-a_{1}b_{2}}{b_{1}},\;\;
 d_{1}=\frac{c_{1}b_{3}-b_{1}c_{2}}{c_{1}}.
 \end{split}
\end{equation}

In Fig.~\ref{fig1}, we represent the double solutions $A_{+}$  and $A_{-}$ of the steady state amplitudes versus $q_{i}$ and $\gamma_{_{X=Y}}$,  where the dissipative parameter $q_{i}$ represents the nonlinear loss related to the self-defocusing Kerr nonlinearity ($q_{r}<0$),  and the parameter $\gamma_{_{X=Y}}$ denotes the symmetric nonlocality degree along the transverse directions. By using the normalized coefficient  $N_{r}=2^{1/2}(4/3)^{3/4}$~\cite{SkarkaPRA2010},  assuming a linear gain $\gamma_{r}$ under anomalous dispersion,  the following parameter values have been used for illustration: $p_{r}=0.579/N_{r}$,  $p_{i}=0.201/N_{r}$, $q_{r}=-110/N_{r}$,  $\gamma_{r}=150/N_{r}$,  $\gamma_{i}=-530/N_{r}$, $k_{0}=2\pi/1.55$.
As stated before~\cite{SkarkaJOA2008}, the stability of the solution $ A_{-} $ is only a prerequisite to obtaining optical light bullets after a spatiotemporal self-organized evolution. It can clearly be seen in Fig.~\ref{fig1} that when the spatial nonlocal self-defocusing Kerr nonlinearity response ($\gamma_{_{X=Y}}$)  and the nonlinear loss ($q_{i}$) increase, the stability zone (yellow area representing $A_{-}$) gets expanded. Hence, the nonlocality and the nonlinear loss enhance optical light bullets localization in this regime.

In Fig.~\ref{fig2}(a) and (b), the stability zones (hatched areas) are shown for the spatial width versus $q_{i}$ and $\gamma_{_{X=Y}}$, respectively, while the same procedure is repeated for the temporal widths $\sigma_{_{X=Y}}$ in Fig.~\ref{fig2}(c) and (d) for the stable solution $A_{-}$ under anomalous dispersion. The set of spatial and temporal widths values derived from these diagrams will be chosen from this stability range for the numerical simulations of the input spatiotemporal pulses.

 Zones of stability for the spatial and temporal chirps, $\vartheta_{_{X=Y}}$  and $\vartheta_{\tau}$, respectively, versus $q_{i}$ and  $\gamma_{_{X=Y}}$, for the stable solution $A_{-}$ and the anomalous dispersion are depicted in Fig.~\ref{fig3}. The set of spatial and temporal chirps derived from these curves will be chosen from this stability range for the numerical simulations of the input spatiotemporal pulses.
\begin{figure}[t]
\centering
\includegraphics[width=3.650in]{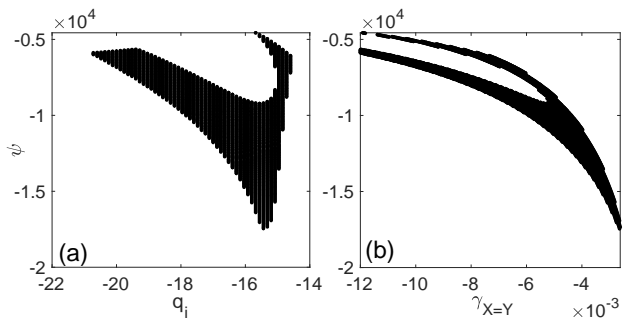}
\caption{Panel (a) shows the stability zone (dark area) for the solution $A_{-}$ in the  $(q_{i}, \psi)-$plane, while panel (b) displays the stability zone in the $(q_{i}, \gamma_{_{X=Y}})-$plane, under anomalous dispersion. Parameter values are those used in Fig.~\ref{fig1}.}
\label{fig4}
\end{figure}

The stability zones for the phase $\psi$ are depicted in Fig.~\ref{fig4}(a) and (b) versus $q_{i}$ and $\gamma_{_{X=Y}}$, respectively, for the stable solution $A_{-}$ under the anomalous dispersion. The set of phase values obtained from these diagrams will be chosen from this stability range for the numerical implementation of the input spatiotemporal pulses.
\begin{figure}[t]
	\centering
	\includegraphics[width=3.0in]{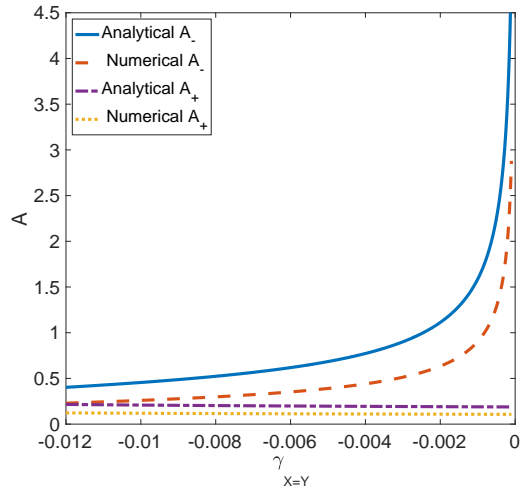}
	\caption{ Analytical and numerical Bifurcation curves of the steady state solutions $A_{-}$ and $A_{+}$ for anomalous dispersion, respectively,  $p_{r}=0.579/N_{r}$ and for the following parameter values: $N_{r}=2^{1/2}(4/3)^{3/4}$,  $p_{i}=0.201/N_{r}$ , $q_{r}=-110/N_{r}$, $q_{i}=-14.7/N_{r}$,  $\gamma_{r}=150/N_{r}$,  $\gamma_{i}=-530/N_{r}$,   and $k_{0}=2\pi/1.55$.}
	\label{fig5}
\end{figure}
\begin{figure}[t]
	\centering
\includegraphics[width=3.0in]{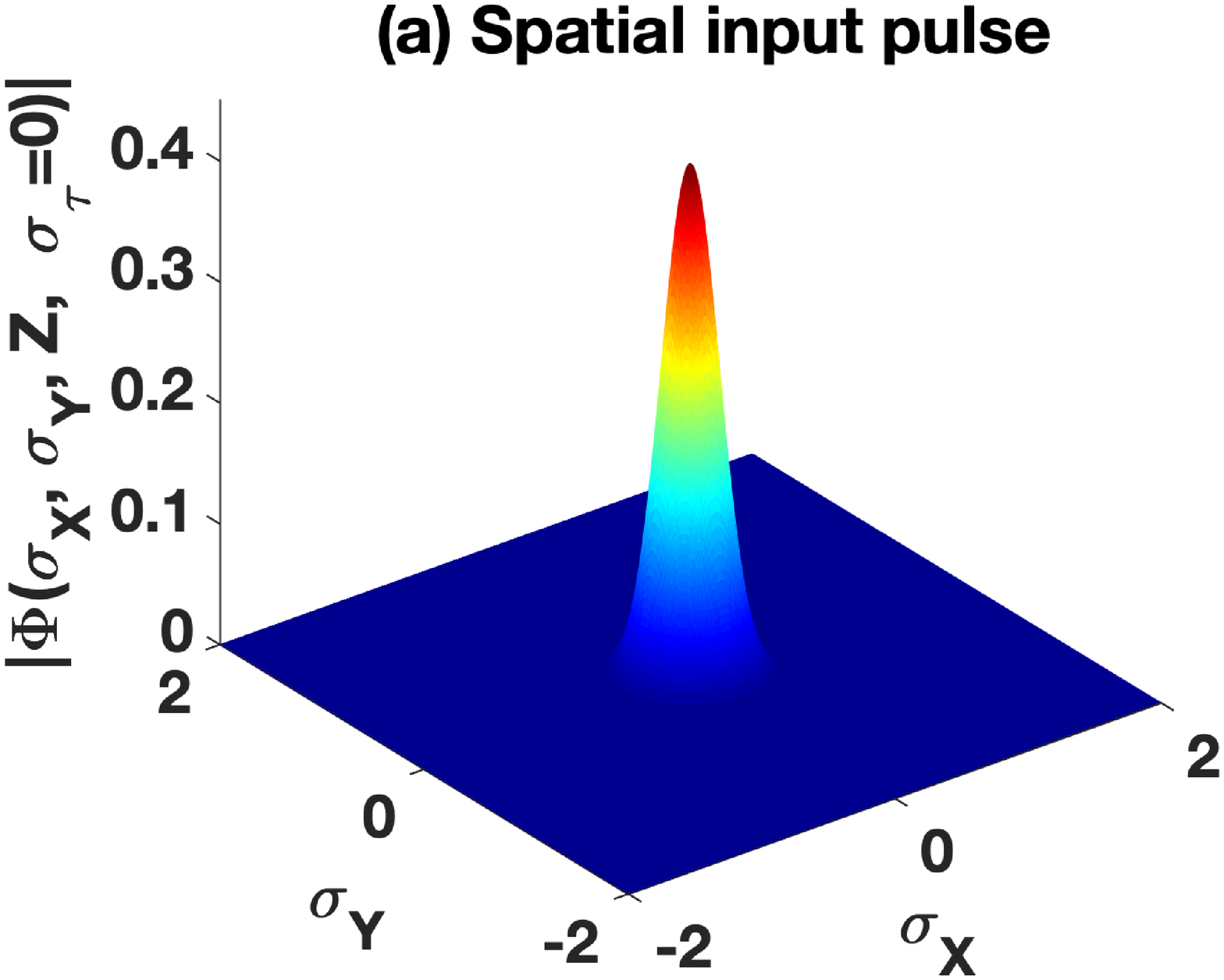}
\includegraphics[width=3.0in]{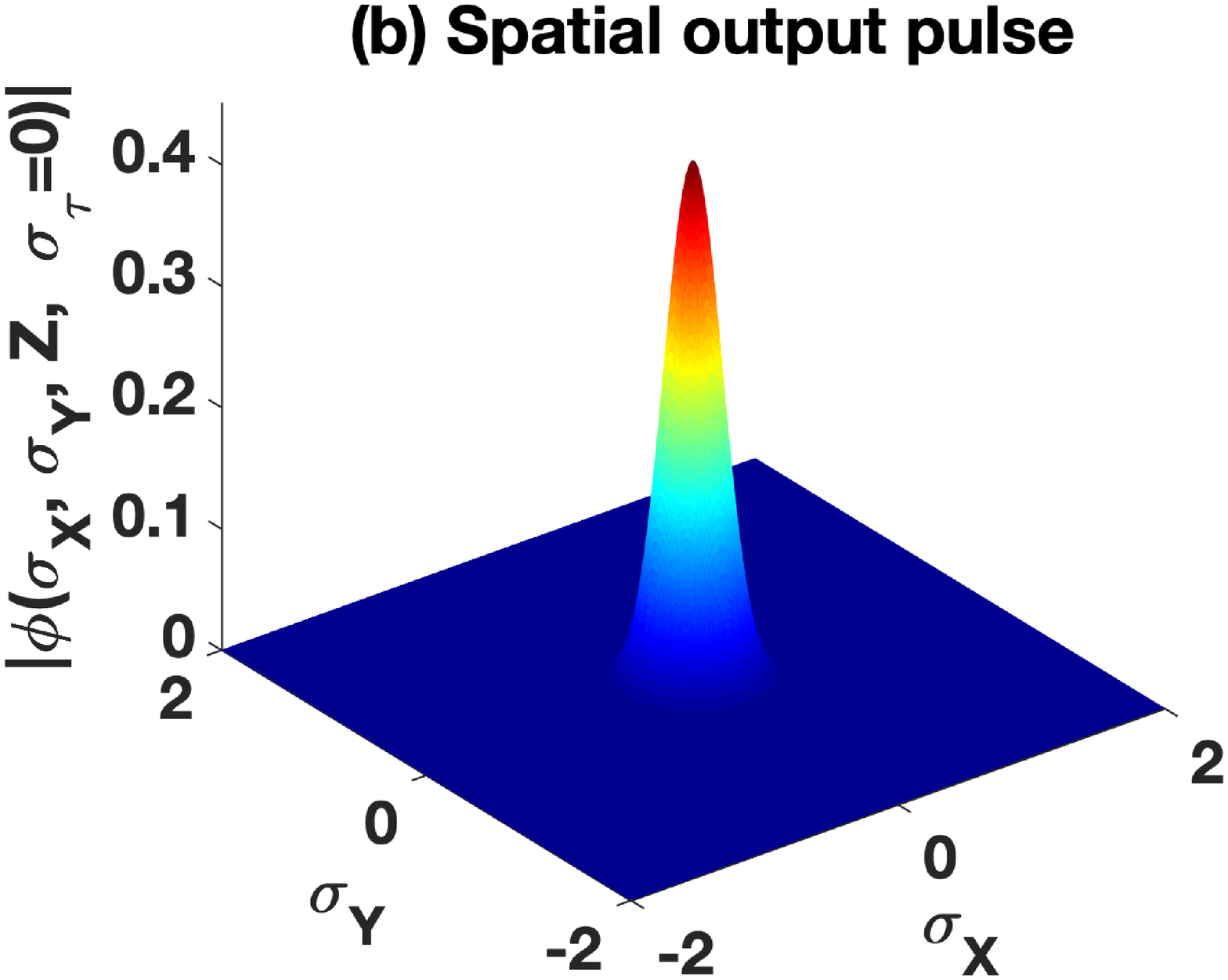}
\caption{  Profiles of the spatial input [panel (a)] and output pulse [panel (b)] for the stable steady state solution $A_{-}$ under anomalous dispersion. The following parameters have been used: $ \gamma_{_{_{X=Y}}}=-0.007$, $ \sigma_{_{X_{0}}}=\sigma_{_{Y_{0}}}=0.2439$, $\vartheta_{X_{0}}=\vartheta_{Y_{0}}=-56.03$, $A_{0-}=0.4225$,   $p_{r}=0.579/N_{r}$, $q_{i}=-14.7/N_{r}$, $p_{i}=0.201/N_{r}$, $q_{r}=-110/N_{r}$,  $\gamma_{r}=150/N_{r}$,  $\gamma_{i}=-530/N_{r}$, $k_{0}=2\pi/1.55$, and $N_{r}=2^{1/2}(4/3)^{3/4}$.}
	\label{fig6}
\end{figure}

\section{Numerical experiments}\label{sec5}

Numerical studies of the evolution of the dissipative light bullet along a doped and weakly nonlocal optical fiber are carried out by means of the fourth-order Runge-Kutta computational method and the split-step Fourier method. The accuracy of numerical experiments is examined by testing different time and space steps. The mesh sizes are chosen as $\Delta X$=$\Delta Y$=0.002 and $\Delta \tau$=0.003. Then, we solve the original (3+1)D nonlocal cubic CGL equation given in Eq.~(\ref{eq23}) via the split-step Fourier method with the longitudinal step size $\Delta Z$=$0.063 \times 10^{-6}$. The following typical optical pulse parameters used in fiber-optic communication systems are adopted~\cite{AgrawalPRA1991,DjokoCNSNS2019,agrawalBook2013}: the wavelength $\lambda=1.55 $ $\mu$m, the linear refractive index $n_{0}=1.45$ cm$^{2}$/W, the nonlinear refractive index $n_{2}=2.7\times 10^{-13}$ cm$^{2}$/W, the group velocity dispersion $\beta_{2}=50$ ps$^{2}$/km, the nonlinear gain $g_{p}=6.8$  W$^{-1}$km$^{-1}$, the pulse width $1.763 T_{0}=400$ fs, the peak power of the incident pulse $P_{0}=9.43$ MW, and the nonlinear parameter $q_{r}$.	  
\begin{figure}[t]
\centering
\includegraphics[width=3.50in]{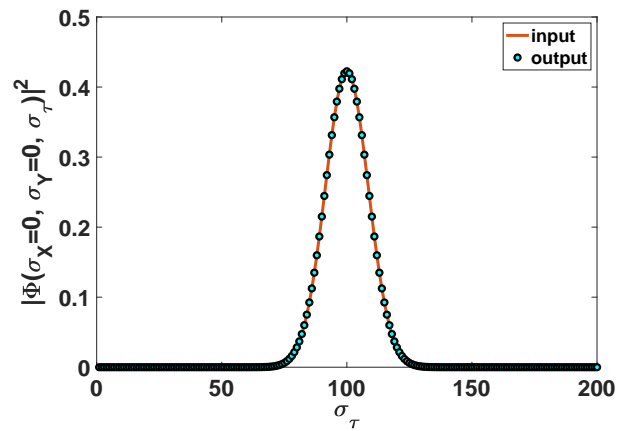}
\caption{Profiles of the temporal input and output pulse for the stable steady state solution $A_{-}$ and for anomalous dispersion. The following parameters have been used: $ \gamma_{_{_{X=Y}}}=-0.007$, $ \sigma_{\tau_{0}}=0.01217$  , $ \vartheta_{\tau_{0}}=572.8$, $A_{0-}=0.4225$,   $p_{r}=0.579/N_{r}$, $q_{i}=-14.7/N_{r}$, $p_{i}=0.201/N_{r}$, $q_{r}=-110/N_{r}$,  $\gamma_{r}=150/N_{r}$,  $\gamma_{i}=-530/N_{r}$, $k_{0}=2\pi/1.55$, and $N_{r}=2^{1/2}(4/3)^{3/4}$.	}
\label{fig7}
\end{figure}
\begin{figure}[t]
\centering
\includegraphics[width=3.0in]{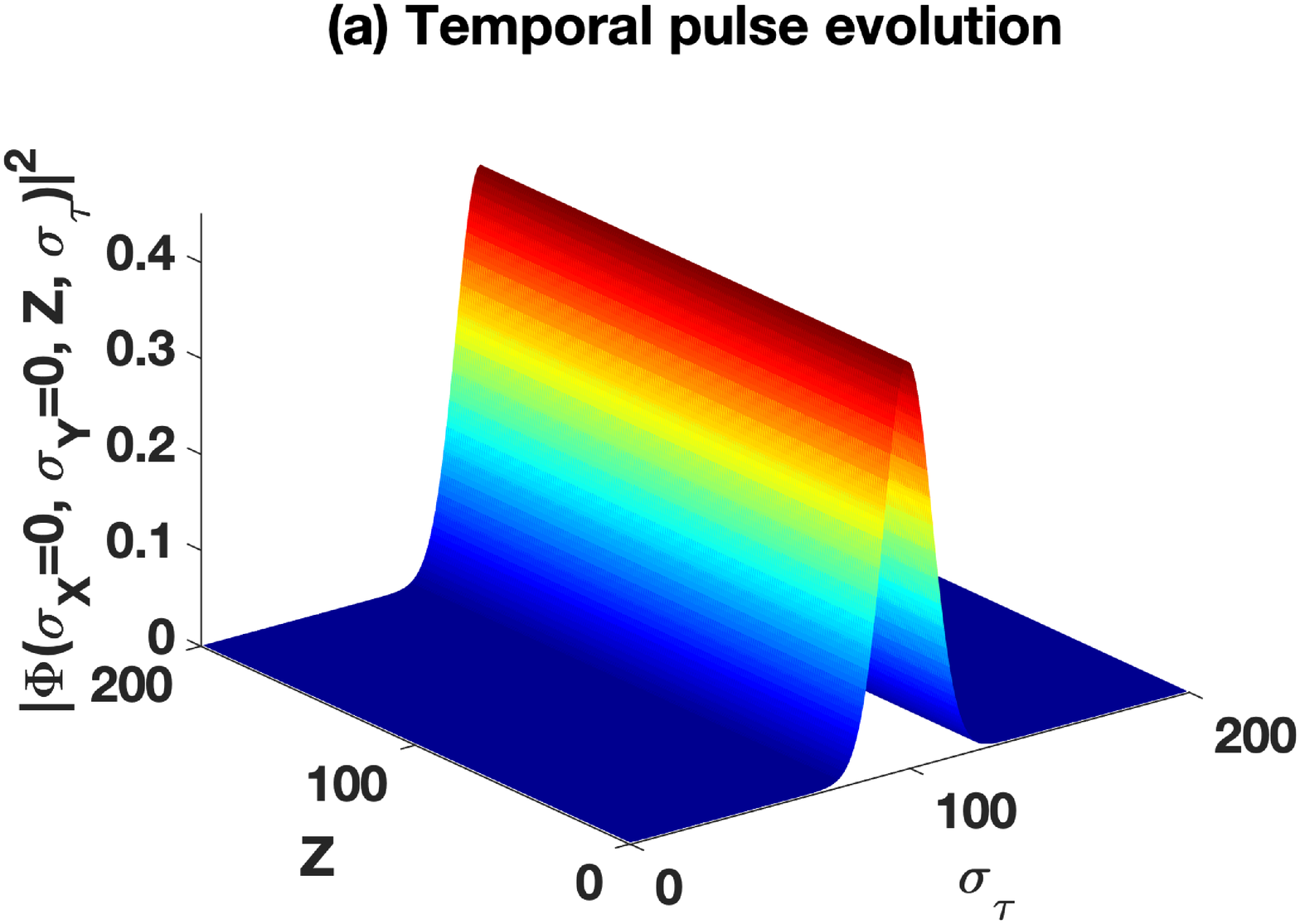}
\includegraphics[width=3.0in]{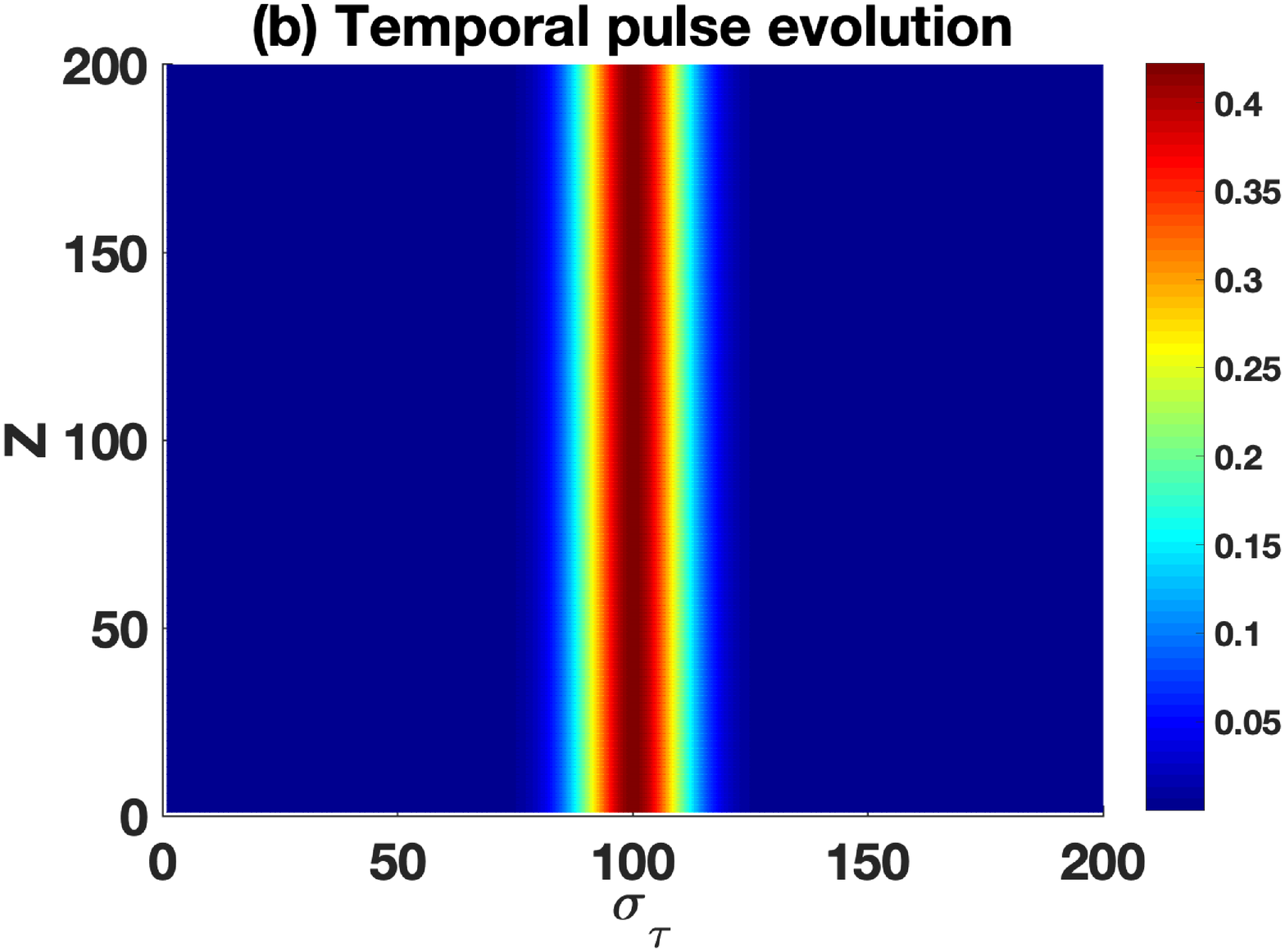}
\caption{Profile of the temporal pulse evolution [panel (a)] and its corresponding density plot [panel(b)] for the stable steady state solution $A_{-}$ under anomalous dispersion. The following parameters have been used: $ \gamma_{_{_{X=Y}}}=-0.007$, $ \sigma_{\tau_{0}}=0.01217$  , $ \vartheta_{\tau_{0}}=572.8$, $A_{0-}=0.4225$,   $p_{r}=0.579/N_{r}$, $q_{i}=-14.7/N_{r}$, $p_{i}=0.201/N_{r}$, $q_{r}=-110/N_{r}$,  $\gamma_{r}=150/N_{r}$,  $\gamma_{i}=-530/N_{r}$, $k_{0}=2\pi/1.55$, and $N_{r}=2^{1/2}(4/3)^{3/4}$. }
\label{fig8}
\end{figure}

From the results of Fig.~\ref{fig5}, we notice that the obtained analytical features corresponding to the steady-state solutions of the amplitudes $A_{-}$ and $A_{+}$ as functions of the spatial nonlocality parameter $\gamma_{_{X=Y}}$, respectively, are a good approximation of the numerically obtained curves. Along the same line, the analytical and numerical solutions of the steady-state $A_{-}$ highlight the upper stable branches. On the contrary, the curves describing the analytical and numerical solutions of the steady-state $A_{+}$ are on the lower unstable branches. For the stable evolution of the self-organized dissipative light bullets represented in numerical simulations, we choose the stable solution $A_{-}$ as an input spatiotemporal pulse.

Fig.~\ref{fig6} shows clearly that the light pulse remains practically constant during its evolution in the spatial domain. We can then notice that the anomalous dispersion, the linear diffraction, the linear gain and the nonlinear diffraction, the spatial nonlocal self-defocusing Kerr nonlinearity response, and the nonlinear loss are well balanced.
	 
In Fig.~\ref{fig7}, the temporal input and output are identical, showing that no loss has been observed.  In the same way, we can notice that the anomalous dispersion, the linear diffraction, the linear gain and the nonlinear diffraction, the spatial nonlocal self-defocusing Kerr nonlinearity response, and the nonlinear loss are well balanced. From the dynamical behaviors depicted in Fig.~\ref{fig8}, we show the evolution of the temporal field profile of the dissipative light bullet intensity distribution in the propagation regime of anomalous dispersion, where the input and output pulses are similar, further confirming our stability predictions, under well-balanced competition from the various involved effects in addition to nonlocality. 
	
\section{Concluding remarks}\label{sec6}

In summary, we have predicted the dissipative light bullets in optical fiber amplifiers for a number of reasons: (i) we have rigorously derived a (3+1)D nonlocal cubic CGL equation valid for the dynamics of the dissipative light bullets in optical fiber amplifiers under the effects fiber dispersion, linear gain, nonlinear loss, fiber nonlinearity, atomic detuning, linear and nonlinear diffractive transverse effects, and nonlocal nonlinear response. (ii) We have also derived eight coupled first-order differential equations of motion of the dissipative light bullet parameters for the nonlocal (3+1)D CGL equation under the interplay between dopants and a spatially weakly nonlocal nonlinear response,  with the help of the variational technique using the Gaussian ansatz function. (iii) We have established a Routh-Hurwitz stability criterion for dissipative spatiotemporal light bullets, where a domain of dissipative parameters for stable steady-state solutions has been found. (iv) We have carried out the direct integration of the proposed nonlocal evolution equation, which allowed us to investigate the evolution of the Gaussian beam along a doped nonlocal optical fiber, showing stable self-organized dissipative spatiotemporal light bullets.
		
Considering the nonlocal CGL equation that we have derived, there are undoubtedly systems in other fields to which they would also apply. For example, Kuramoto~\cite{KuramotoPTP1995} has proposed the nonlocal CGL equation for populations of biologically oscillating cells secreting substances whose rapid diffusion mediates the cell-cell interaction. It was found that under certain conditions, the correlations and fluctuations obey a power law similar to the one in the fully developed Navier-Stokes turbulence. Also, effective nonlocality in coupling may become relevant when the reaction-diffusion system involves three or more chemical components. Thus, Tanaka and Kuramoto~\cite{TanakaPRE2003} have proposed the nonlocal CGL equation as a reduced form of a universal class of reaction-diffusion systems near the Hopf bifurcation. In this context, novel dynamical states have been predicted, such as multi-affine chemical turbulence~\cite{KuramotoPRL1998} and chimera states~\cite{AbramsPRL2004}. In addition, the nonlocal CGL equation has been used extensively to study electrochemical turbulence for electrochemical systems with migration coupling~\cite{GarciaPRL2008,VarelaPRL2005}. Indeed, oscillatory electrochemical systems can be considered active distributed media and are mathematically described by a set of coupled partial differential equations. They only differ from the reaction-diffusion system in the spatial coupling term of the electric potential drop across the electrode/electrolyte interface. The spatial coupling in electrochemical systems is nonlocal. Some coherent structures have been found in the nonlocal CGL equation in the turbulent regime, which includes standing waves and robust heteroclinic orbits between fixed points or limit cycles~\cite{GarciaPRE2008}. We believe that such studies can be extended to the proposed model by studying their azimuthal manifestation on spherical surfaces, for example. Investigations in that direction are ongoing and will be published elsewhere.

\begin{acknowledgements}

The work by CBT is supported by the Botswana International University of Science and Technology under the grant {\bf DVC/RDI/2/1/16I (25)}. CBT thanks the Kavli Institute for Theoretical Physics (KITP), the University of California Santa Barbara (USA),  where this work was supported in part by the National Science Foundation Grant no.{\bf NSF PHY-1748958}, NIH Grant no.{\bf R25GM067110}, and the Gordon and Betty Moore Foundation Grant no.{\bf 2919.01}.

\end{acknowledgements}
\begin{widetext}				
\section*{APPENDIX: Partial derivatives of the functions $ F_{A}$, $ F_{\sigma}$, $ F_{\vartheta}$, $ F_{\sigma_{\tau}}$, $ F_{\vartheta_{\tau}}$}
\appendix*		
\begin{equation*}
\begin{split}
\frac{\partial F_{A}}{\partial A }&=-\zeta_{1}k_{0}\vartheta - \zeta_{1}k_{0}\vartheta-\zeta_{2}\vartheta_{\tau}- (7\zeta_{3}b_{1})/2 -(231A^2\sqrt{2}c_{0}\zeta_{4})/16+a_{1}\zeta_{2}(35k_{0}^2\vartheta_{\tau}^2\sigma_{\tau}^2+ 156/\sigma_{\tau}^2)/8 \\
&+(3A^2c_{_{X=Y}}\zeta_{_{X=Y}}\sqrt{2}(75k_{0}^2\vartheta^2\sigma^2+ 996/\sigma^2))/128+(3A^2c_{_{X=Y}}\zeta_{_{X=Y}}\sqrt{2}(75k_{0}^2\vartheta^2\sigma^2+ 996/\sigma^2))/12,\\
\frac{\partial F_{\sigma}}{\partial A}&=(15A\sqrt{2}c_{0}\zeta_{4}\sigma)/4+(A\sqrt{2}c_{_{X=Y}}\zeta_{_{X=Y}}(13k_{0}^2\vartheta^2\sigma^2 - 252/\sigma^2)\sigma)/32 \\
&+(15A\sqrt{2}c_{_{X=Y}}\zeta_{_{X=Y}}(-k_{0}^2\vartheta^2\sigma^2 - 12/\sigma^2)\sigma)/32,\\
\frac{\partial F_{\sigma_{\tau}}}{\partial A }&=(15A\sqrt{2}c_{0}\zeta_{4}\sigma_{\tau})/4 +(15A\sqrt{2}c_{_{X=Y}}\zeta_{_{X=Y}}(-k_{0}^2\vartheta^2\sigma^2 - 12/\sigma^2)\sigma_{\tau})/32 \\
&+(15A\sqrt{2}c_{_{X=Y}}\zeta_{_{X=Y}}(-k_{0}^2\vartheta^2\sigma^2 - 12/\sigma^2)\sigma_{\tau})/32,
\end{split}
\end{equation*}
\begin{equation*}
\begin{split}
\frac{\partial F_{\vartheta}}{\partial A }&=2A\sqrt{2}\zeta_{4}/(k_{0}\sigma^2) +(A\sqrt{2}c_{_{X=Y}}\zeta_{_{X=Y}})(24/\sigma^2+ 5k_{0}\vartheta)/(2k_{0}\sigma^2)+(A\sqrt{2}c_{_{X=Y}}\zeta_{_{X=Y}})(8/\sigma^2 -k_{0}\vartheta)/(2k_{0}\sigma^2),\\	
\frac{\partial F_{\vartheta_{\tau}}}{\partial A }&=2A\sqrt{2}\zeta_{4}/(k_{0}\sigma_{\tau}^2) +(A\sqrt{2}c_{_{X=Y}}\zeta_{_{X=Y}} )(8/\sigma^2 -k_{0}\vartheta)/(2k_{0}\sigma_{\tau}^2)+(A\sqrt{2}c_{_{X=Y}}\zeta_{_{X=Y}})(8/\sigma^2 -k_{0}\vartheta)/(2k_{0}\sigma_{\tau}^2),\\
\frac{\partial F_{A}}{\partial \sigma }&=A^3c_{_{X=Y}}\zeta_{_{X=Y}}\sqrt{2}(150k_{0}^2\vartheta^2\sigma - 1992/\sigma^3)/128,
\end{split}
\end{equation*}
\begin{equation*}
\begin{split}
\frac{\partial F_{\sigma}}{\partial \sigma } &=2\zeta_{1}k_{0}\vartheta + \zeta_{3}b_{1} +(15A^2\sqrt{2}c_{0}\zeta_{4})/8 + a_{1}\zeta_{2}(-7k_{0}^2\vartheta_{\tau}^2\sigma_{\tau}^2 -28/\sigma_{\tau}^2)/4+A^2\sqrt{2} c_{_{X=Y}}\zeta_{_{X=Y}}(26k_{0}^2\vartheta^2\sigma +504/\sigma^3)\sigma/64 \\
&+A^2\sqrt{2}c_{_{X=Y}}\zeta_{_{X=Y}}(13k_{0}^2\vartheta^2\sigma^2-252/\sigma^2)/64+(15A^2\sqrt{2}c_{_{X=Y}}\zeta_{_{X=Y}}(-k_{0}^2\vartheta^2\sigma^2 -12/\sigma^2))/64,\\
\frac{\partial F_{\sigma_{\tau}}}{\partial \sigma}&=(15A^2\sqrt{2}c_{_{X=Y}}\zeta_{_{X=Y}}(-2k_{0}^2\vartheta^2\sigma+24/\sigma^3)\sigma_{\tau})/64,\\
\frac{\partial F_{\vartheta}}{\partial \sigma } &= -2A^2\sqrt{2}\zeta_{4}/(k_{0}\sigma^3)-32\zeta_{1}/(k_{0}\sigma^5)-c_{_{X=Y}}\zeta_{_{X=Y}}A^2\sqrt{2}(24/\sigma^2 +5k_{0}\vartheta)/(2k_{0}\sigma^3) - 12c_{_{X=Y}}\zeta_{_{X=Y}}A^2\sqrt{2}/(k_{0}\sigma^5)\\
&- A^2\sqrt{2}c_{_{X=Y}}\zeta_{_{X=Y}}(8/\sigma^2- k_{0}\vartheta)/(2k_{0}\sigma^3),
\end{split}
\end{equation*}
\begin{equation*}
\begin{split}
\frac{\partial F_{\vartheta_{\tau}}}{\partial \sigma } &= -4c_{_{X=Y}}\zeta_{_{X=Y}}A^2\sqrt{2}/(\sigma^3k_{0}\sigma_{\tau}^2),\;\;
\frac{\partial F_{A}}{\partial \sigma_{\tau}}= Aa_{1}\zeta_{2}(70k_{0}^2\vartheta_{\tau}^2\sigma_{\tau}- 312/\sigma_{\tau}^3)/8,\\
\frac{\partial F_{\sigma}}{\partial \sigma_{\tau} }&= a_{1}\zeta_{2}\sigma(-14k_{0}^2\vartheta_{\tau}^2\sigma_{\tau} + 56/\sigma_{\tau}^3)/4,\\
\frac{\partial F_{\sigma_{\tau}}}{\partial \sigma_{\tau} }&=2\zeta_{2}\vartheta_{\tau}+(15A^2\sqrt{2}c_{0}\zeta_{4})/8+\zeta_{3}b_{1}
+a_{1}\zeta_{2}(-5k_{0}^2\vartheta_{\tau}^2\sigma_{\tau}^2 - 36/\sigma_{\tau}^2)/4 +a_{1}\zeta_{2}\sigma_{\tau}(10k_{0}^2\vartheta_{\tau}^2\sigma_{\tau}+ 72/\sigma_{\tau}^3)/4 \\
&+(15A^2\sqrt{2}c_{_{X=Y}}\zeta_{_{X=Y}}(-k_{0}^2\vartheta^2\sigma^2 - 12/\sigma^2))/32,
\end{split}
\end{equation*}
\begin{equation*}
\begin{split}
\frac{\partial F_{\vartheta}}{\partial \sigma_{\tau} }&= 0 ,\;\;
\frac{\partial F_{\vartheta_{\tau}}}{\partial \sigma_{\tau} }= -2A^2\sqrt{2}\zeta_{4}/(k_{0}\sigma_{\tau}^3)-16\zeta_{2}a_{1}\vartheta_{\tau}/\sigma_{\tau}^3+ 32\zeta_{2}/(k_{0}\sigma_{\tau}^5)-\zeta_{_{X=Y}}c_{_{X=Y}}A^2\sqrt{2}(8/\sigma^2-k_{0}\vartheta)/(k_{0}\sigma_{\tau}^3),\\
\frac{\partial F_{A}}{\partial \vartheta}&= -A\zeta_{1}k_{0} +(75A^3c_{_{X=Y}}\zeta_{_{X=Y}}\sqrt{2}k_{0}^2\vartheta\sigma^2)/64,\\
\frac{\partial F_{\sigma}}{\partial \vartheta }&= 2\zeta_{1}k_{0}\sigma+(13c_{_{X=Y}}\zeta_{_{X=Y}}A^2\sqrt{2}k_{0}^2\vartheta\sigma^3)/32,\\
\frac{\partial F_{\sigma_{\tau}}}{\partial \vartheta }&=-(15A^2\sqrt{2}c_{_{X=Y}}\zeta_{_{X=Y}}k_{0}^2\vartheta\sigma^2\sigma_{\tau})/32,
\end{split}
\end{equation*}
\begin{equation}
\begin{split}
\frac{\partial F_{\vartheta}}{\partial \vartheta }&= -4\zeta_{1}k_{0}\vartheta +(50\zeta_{_{X=Y}}c_{_{X=Y}}A^2\sqrt{2})/(9\sigma^2),\;\;
\frac{\partial F_{\vartheta_{\tau}}}{\partial \vartheta }= -(\zeta_{_{X=Y}}c_{_{X=Y}}A^2\sqrt{2})/(4\sigma_{\tau}^2),\\
\frac{\partial F_{A}}{\partial \vartheta_{\tau}}&= A\zeta_{2} + (35/4) Aa_{1}\zeta_{2}k_{0}^2\vartheta_{\tau}\sigma_{\tau}^2,\\
\frac{\partial F_{\sigma}}{\partial \vartheta_{\tau} }&= -(7a_{1}\zeta_{2}\sigma k_{0}^2\vartheta_{\tau}\sigma_{\tau}^2)/2,\;\;
\frac{\partial F_{\sigma_{\tau}}}{\partial \vartheta_{\tau} }= 2\zeta_{2}\sigma_{\tau}-5/2a_{1}\zeta_{2}\sigma_{\tau}^3k_{0}^2\vartheta_{\tau},\\
\frac{\partial F_{\vartheta}}{\partial \vartheta_{\tau} }&= 0,\;\;
\frac{\partial F_{\vartheta_{\tau}}}{\partial \vartheta_{\tau} }= 8\zeta_{2}a_{1}/\sigma_{\tau}^2-4\zeta_{2}k_{0}\vartheta_{\tau}.
\end{split}
\end{equation}
\end{widetext}

\end{document}